\documentclass[final,5p,times,twocolumn]{elsarticle}

\usepackage[utf8]{inputenc}
\usepackage[T1]{fontenc}
\usepackage{amsmath,amssymb,bm,mathtools,amsfonts}
\usepackage{graphicx}
\usepackage{booktabs}
\usepackage{array}
\usepackage{tabularx}
\usepackage{adjustbox}
\usepackage{subcaption}
\usepackage[hidelinks]{hyperref}
\usepackage{xcolor}
\usepackage{textcomp}
\usepackage{mathrsfs}
\usepackage{dblfloatfix}
\usepackage{tikz}
\usepackage{microtype}
\usetikzlibrary{positioning,arrows.meta}
\biboptions{numbers,sort&compress}

\newcolumntype{Y}{>{\raggedright\arraybackslash}X}
\newcommand{\vect}[1]{\boldsymbol{#1}}
\newcommand{\dd}{\mathrm{d}}
\newcommand{\norm}[1]{\left\lVert #1 \right\rVert}
\DeclareMathOperator{\axl}{axl}

\begin{document}

\begin{frontmatter}

\title{Contact-resolved deployment of the Contour Neurovascular System in patient-specific intracranial aneurysms}

\author[unibw]{Ratnadeep Pramanik\corref{cor1}}
\ead{ratnadeep.pramanik@unibw.de}

\author[kiel1]{Fina Gie\ss ler}
\author[unibw]{Martin Frank}
\author[unibw]{Ivo Steinbrecher}
\author[unibw]{Matthias Mayr}
\author[kiel2]{Naomi Larsen}
\author[kiel1]{Sylvia Saalfeld}
\author[unibw]{Alexander Popp}

\cortext[cor1]{Corresponding author}

\address[unibw]{Institute for Mathematics and Computer-Based Simulation, University of the Bundeswehr Munich, 85577 Neubiberg, Germany}
\address[kiel1]{Institute for Medical Informatics and Statistics, Kiel University and University Hospital Schleswig-Holstein, 24105 Kiel, Germany}
\address[kiel2]{Department of Radiology and Neuroradiology, University Hospital Schleswig-Holstein, Kiel University, 24105 Kiel, Germany}

\begin{abstract}
While intrasaccular flow disruptors are increasingly used to treat wide-necked intracranial aneurysms (IAs), many patient-specific computational workflows prescribe a pre-seated device geometry and omit deployment mechanics. This simplification is particularly restrictive for the Contour Neurovascular System (CNS), for which neck coverage, wall apposition, and post-contact motion depend on the deployment process. To address this, we present a contact-resolved finite-element framework that models CNS deployment within patient-specific IAs. We represent the device as a dual-layer interwoven Nitinol braid using geometrically exact beam models, and the aneurysm wall as a deformable hyperelastic shell. Frictional contact governs wire-wire and wire-wall interactions during staged release. The framework is demonstrated across three patient-specific anatomies. In the reference anatomy, the seated configuration is sensitive to the assumed device-wall tangential friction and to release depth relative to the aneurysm neck plane. Frictionless wall contact permits pronounced post-contact sliding, whereas finite tangential friction strongly suppresses residual pole motion. Release depth alters both the onset of wall engagement and the subsequent deployment path. Geometric placement approaches that prescribe the implanted configuration cannot recover this contact history or the associated wall-supported state. Our framework provides a mechanics-based route to deployed geometries for downstream hemodynamic, fluid-structure interaction, and mechanobiological analyses.
\end{abstract}


\end{frontmatter}


\section{Introduction}
\label{sec:introduction}

Intracranial aneurysms may remain clinically silent until rupture, after which aneurysmal subarachnoid hemorrhage carries high morbidity and mortality \cite{etminan2016unruptured,backes2016patient,brisman2006cerebral,texakalidis2019aneurysm}. Therefore, treatment decisions depend not only on aneurysm geometry but also on whether the expected rupture risk justifies the procedural risk. Endovascular therapy has become central to that decision because it can treat complex cerebrovascular anatomy with less invasiveness than open surgery \cite{lauzier2023review,fatania2022comprehensive}.

Current endovascular options include stand-alone coiling, stent-assisted coiling, flow diverters, and intrasaccular flow-disruption devices \cite{starke2015endovascular,briganti2015endovascular,heiferman2024new,shao2024intrasaccular}. These technologies differ in where they act. Coils occupy sac volume and increase hydraulic resistance, but recurrence remains relevant in wide-neck lesions \cite{goubergrits2014hemodynamic,maroufi2024comparison}. Intraluminal flow diverters are deployed within the parent artery and redirect blood flow away from the aneurysm through a low-porosity, self-expanding tubular wire mesh \cite{dholakia2017hemodynamics,wang2016flow,jing2016hemodynamic}. Intrasaccular devices such as Woven EndoBridge (WEB) and Contour Neurovascular System (CNS) act at or near the ostium, where stable neck engagement is the central mechanical requirement \cite{muskens2017woven,mantilla2026woven,hecker2023comparison,wodarg2025embolization}.

This mechanical distinction directly dictates the corresponding computational modeling strategy. While the dominant interface for a tubular flow diverter lies along the parent-vessel lumen, an intrasaccular neck-covering implant interacts primarily with the ostium and the neck-adjacent wall. Consequently, key parameters, such as neck coverage, uncovered inflow gaps, pole position, local compaction, wall apposition, and migration tendencies, are fundamentally structural outcomes of the initial deployment process; only secondarily do they serve as downstream hemodynamic descriptors. For the CNS, clinical studies have reported encouraging occlusion performance alongside instances of deformation, migration, lateral impression, inversion, and recurrence in selected cases~\cite{larsen2026morphological,akhunbay2020endovascular,bhogal2021contour,griessenauer2025contour,koc2025delayed}. Longitudinal imaging has further demonstrated that both the aneurysm morphology and the implanted CNS configuration can actively evolve during follow-up~\cite{spitz2026morphological}. Recent cohort analyses have linked these post-treatment shape modifications directly to pre-treatment geometric and flow characteristics, demonstrating that device-associated morphological shifts generate significant variations in aneurysmal hemodynamics that heavily modulate the anticipated treatment effect~\cite{bernovskis2026hemodynamic,giessler2026clinical}. These observations reinforce the necessity of treating the deployed configuration as a mechanically resolved state rather than a simplified, prescribed geometry.

Device sizing and positioning link nominal implant design to the mechanically realized deployment state. For intrasaccular devices, undersizing can leave portions of the ostium uncovered, whereas excessive confinement can alter device shape, compaction, and wall support. Moreover, delayed displacement or inversion may compromise an initially favorable interface and reopen preferential inflow pathways \cite{zhuo2025intrasaccular,bellanger2024reversed}. These observations reinforce that the clinically relevant configuration is not the unconstrained device geometry, but the wall-supported state established after release.

Nevertheless, several computational workflows start by prescribing the implanted geometry \cite{pramanik2026braided}. Rapid virtual placement or image-based reconstruction is often followed by computational fluid dynamics (CFD) to estimate post-treatment inflow, velocity, or wall shear stress \cite{lyu2024treatment,korte2024vitro,korte2025analysis,spitz2024assessment}. Other studies use porous or heterogeneous surrogates to approximate the hydraulic effect of dense braided structures without resolving the deployment mechanics \cite{frank2024numerical,abdehkakha2021cerebral,berod2022heterogeneous,reymond2025novel}. These approaches are useful for screening and flow analysis, but they do not answer a structural question that is decisive for CNS: why does the same device settle into one seated shape in one aneurysm and a different shape in another? Consequently, the flow field computed on a prescribed post-deployment geometry inherits any mechanical error already present in that geometry \cite{berg2019review,campo2015review,boniforti2024endovascular}.

The second gap concerns braid mechanics. Studies of braided stents show that wire count, braid angle, crossover treatment, and tangential interaction assumptions affect stiffness, compaction, and release behavior \cite{kim2008mechanical,de2009virtual,kelly2019comparison,mckenna2021finite,zaccaria2021analytical,abdollahi2024virtual}. Kinematically tied crossover models constrain the intersecting wires to undergo identical relative translations and rotations at each crossover. They therefore suppress tangential slip and changes in the local crossing angle, which can make a braided device artificially stiff. Contact-resolved crossovers instead permit local sliding, rotation, and rearrangement of neighboring wires. The resulting nonlinearity and history dependence arise from the connected braid network undergoing large deformation, evolving contact, and frictional slip. These modeling choices determine whether the final pore geometry and neck-level support are mechanically plausible.

A third gap concerns procedural deployment conditions. For an ostium-covering device, the final seated state can depend on the nominal device size, release depth relative to the neck plane, phase alignment of the braid, and tangential friction after wall contact \cite{pravdivtseva2025effect}. A model (based on fast placement strategies) that prescribes the final geometry cannot expose that dependence. More generally, recent finite-element studies of cardiovascular stent implantation have demonstrated that resolving the complete crimping and deployment sequence, including large deformation and evolving device–tissue contact, is essential for predicting the mechanically realized post-deployment state \cite{tragoudas2026computational}.

In the present work, CNS deployment is modeled as a mixed-dimensional, large-deformation contact problem solved using explicit time integration. This is a mechanically plausible and computationally efficient reduction: the slender Nitinol wires are represented by geometrically exact one-dimensional beams, while a deformable shell surface represents the aneurysm wall. A fully three-dimensional solid discretization of all 144 wires would be possible in principle, but would be substantially more computationally expensive. The present mixed-dimensional reduction is adopted to make wire-resolved patient-specific deployment analyses tractable. Their interaction must account for normal impenetrability, tangential slip resistance, staged release, and the finite radius offset between beam centerlines and wall surfaces. Recent work on mixed-dimensional coupling has shown that offset-aware beam-surface interaction is needed to preserve frame invariance and angular momentum balance \cite{steinbrecher2025consistent,popp2018state}. Variationally consistent beam-beam coupling provides a compatible route for wire-wire interaction in geometrically exact beam formulations \cite{steinbrecher2026pointcoupling}. Mixed-dimensional approaches for vascular device placement have recently been presented \cite{datz2025patient,frank2026mechanical,datz2026influence}.

This study treats the CNS implant as a deployable structural system, not as a porous cap pasted onto the neck plane. The device is represented as a dual-layer interwoven braid discretized by geometrically exact beams. The aneurysm is modeled as a deformable hyperelastic shell. Wire-wire and wire-wall interactions are resolved via frictional contact, and the implanted state is achieved via staged release. The four sequential deployment stages are: (i) a 1 s initial stabilization of the crimped CNS within the catheter, (ii) a 4 s advancement of the catheter-device system into the aneurysm dome, (iii) a 20 s unsheathing and expansion phase where the device unfurls from the catheter to engage the aneurysm wall, during which a temporary Dirichlet boundary condition keeps the pole fixed before its release, capturing highly nonlinear fluctuations in wire-to-wire and wire-to-wall contact area before achieving geometric and mechanical locking, and (iv) a prolonged 35 s seating stage deliberately introduced to monitor late-time structural stability and ensure the complete suppression of residual sliding or migration. The resulting pole motion, contact-area growth, rim coverage, and uncovered neck gaps are structural outputs. The objective is to establish a mechanics-based deployment workflow that can support later hemodynamic analysis, fluid-structure interaction (FSI), and deployment-informed virtual treatment planning.

\section{Models and workflows}
\label{sec:models_workflow}

\begin{figure*}[t!]
\centering
\begin{subfigure}[b]{0.325\textwidth}
\centering
\begin{tikzpicture}
    \node[anchor=south west, inner sep=0] (image) at (0,0) {%
        \adjustbox{trim=250pt 0pt 250pt 0pt, clip, width=\textwidth}{\includegraphics{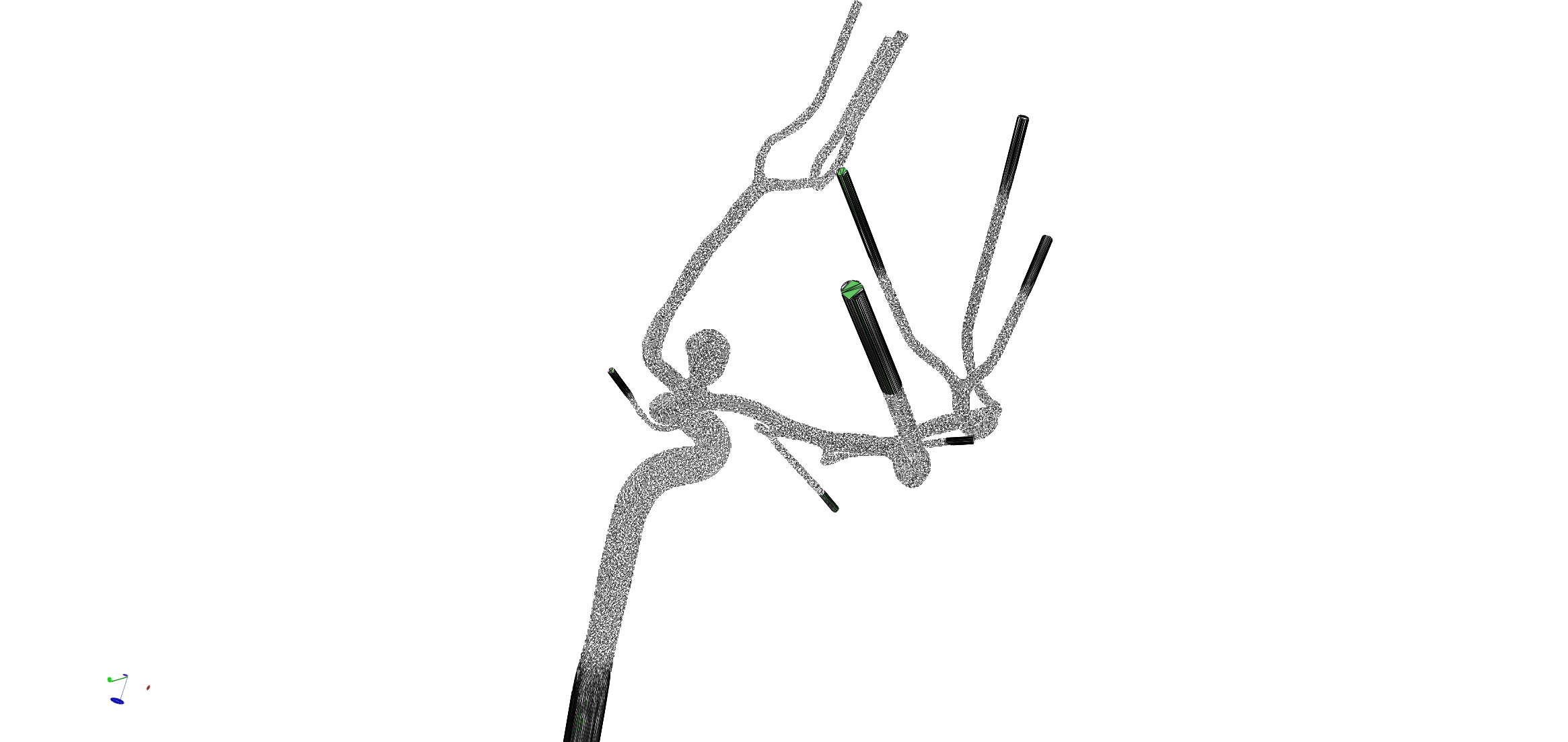}}%
    };
    \draw[red, ultra thick, dotted] ([shift={(-0.62cm,0.15cm)}]image.center) circle (0.4cm);
\end{tikzpicture}
\caption{Segmented lumen for Case I}
\end{subfigure}
\hfill
\begin{subfigure}[b]{0.325\textwidth}
\centering
\begin{tikzpicture}
    \node[anchor=south west, inner sep=0] (image) at (0,0) {%
        \adjustbox{trim=100pt 0pt 100pt 0pt, clip, width=\textwidth}{\includegraphics{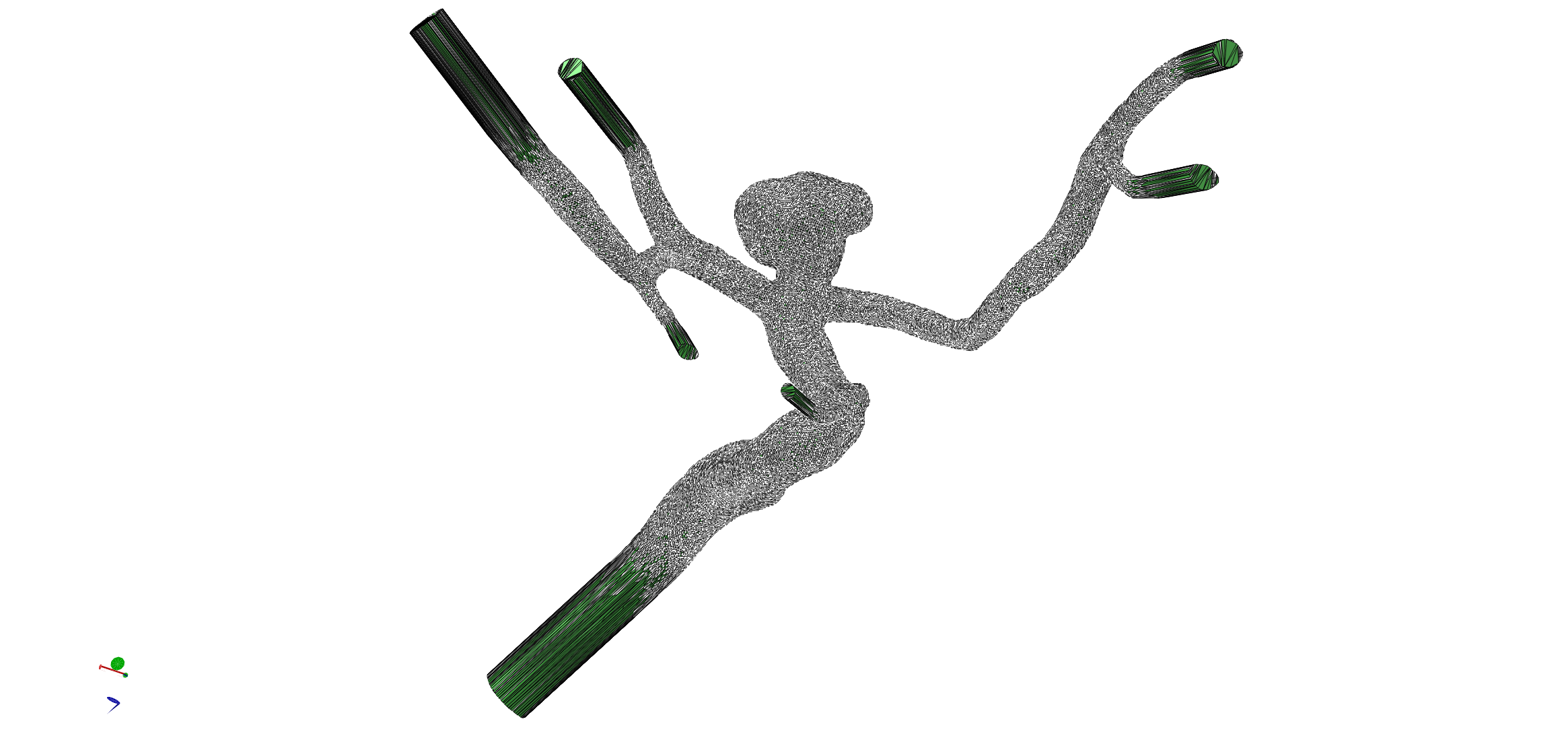}}%
    };
    \draw[red, ultra thick, dotted] ([shift={(0.1cm,0.7cm)}]image.center) circle (0.5cm);
\end{tikzpicture}
\caption{Segmented lumen for Case II}
\end{subfigure}
\hfill
\begin{subfigure}[b]{0.325\textwidth}
\centering
\begin{tikzpicture}
    \node[anchor=south west, inner sep=0] (image) at (0,0) {%
        \adjustbox{trim=250pt 0pt 250pt 0pt, clip, width=\textwidth}{\includegraphics{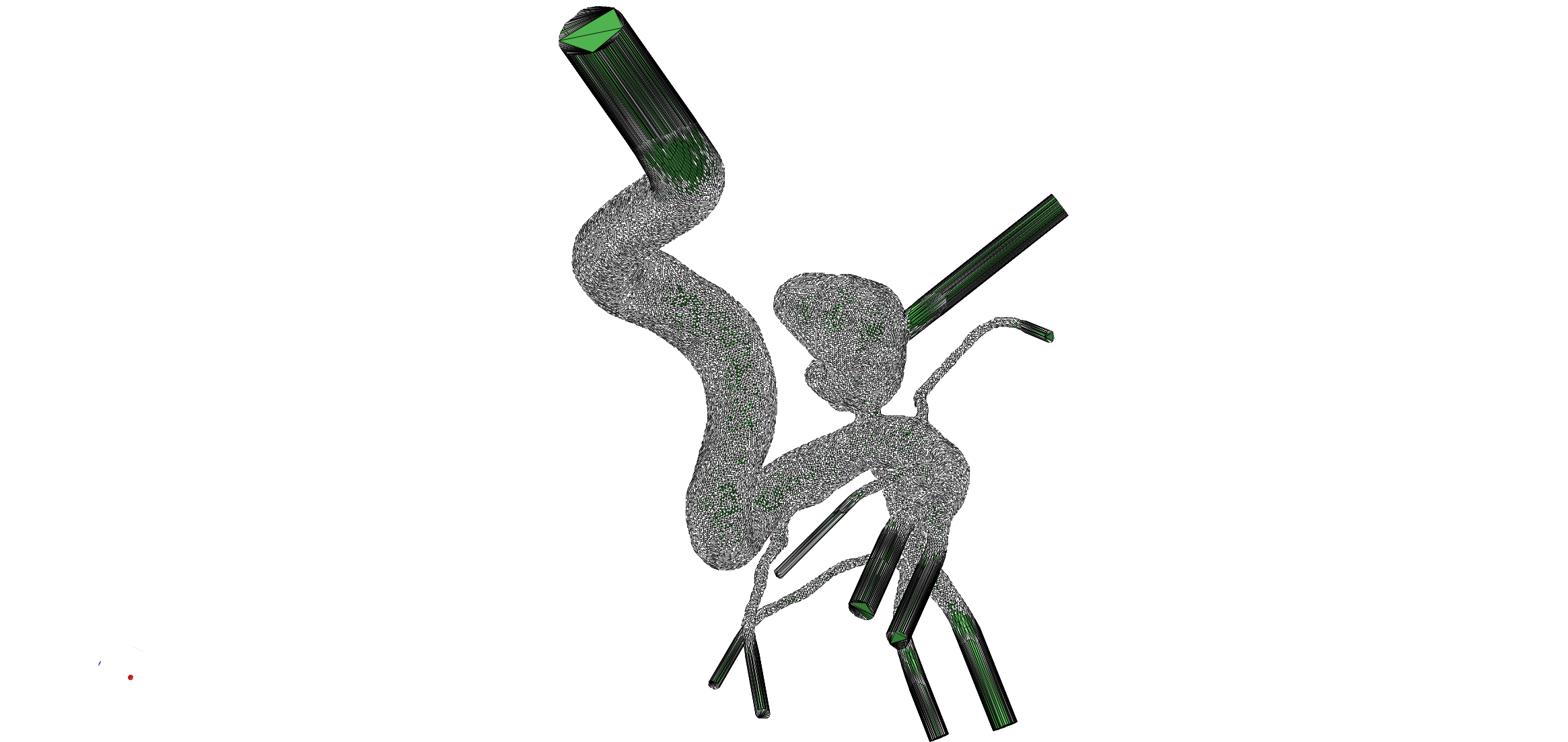}}%
    };
    \draw[red, ultra thick, dotted] ([shift={(0.5cm,0.3cm)}]image.center) circle (0.8cm);
\end{tikzpicture}
\caption{Segmented lumen for Case III}
\end{subfigure}
\caption{Patient-specific vascular geometries reconstructed from pre-treatment 3D rotational angiography (3D-RA). The red dotted circles highlight the targeted intracranial aneurysm regions embedded within the broader parent vessel network prior to structural domain isolation.}
\label{fig:segmented_aneurysms}
\end{figure*}
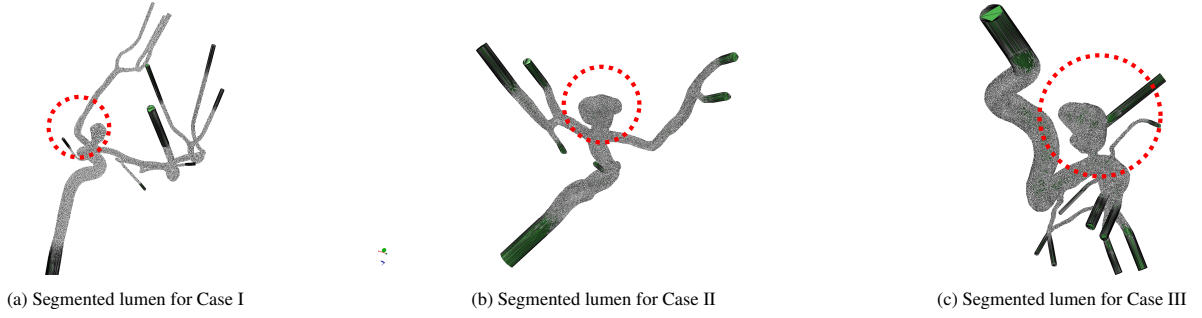

\subsection{Patient-specific anatomies}
\label{subsec:anatomies}

The study uses imaging data from three anonymous patients treated with a 7~mm CNS between 12/2018 and 11/2020. Pre-treatment anatomy is obtained from three-dimensional rotational angiography (3D-RA) acquired at the University Hospital Schleswig-Holstein on a biplane angiography system (Azurion, Philips, Amsterdam, The Netherlands) with a reconstructed voxel size of 0.16~mm. Post-treatment and follow-up data are acquired by computed tomography angiography (CTA). Follow-up ranges from 367 to 1274~days.

\begin{figure*}[htpb!]
\centering
\begin{subfigure}[b]{0.325\textwidth}
\centering
\adjustbox{width=1.4\textwidth, center}{\includegraphics{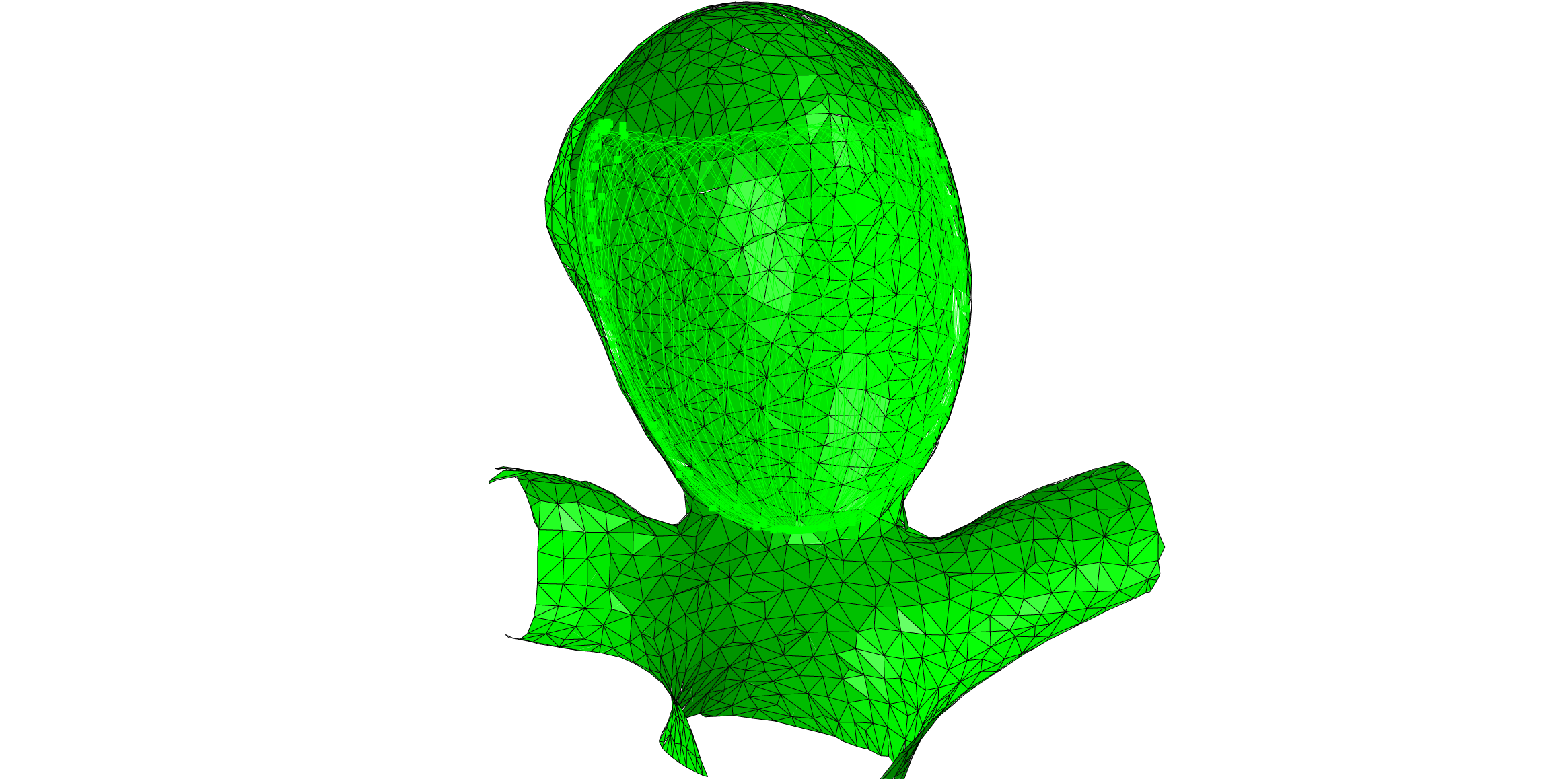}}
\caption{Shell mesh for Case I ($3,717$ elements)}
\label{fig:case_1}
\end{subfigure}
\hfill
\begin{subfigure}[b]{0.325\textwidth}
\centering
\adjustbox{width=1.4\textwidth, center}{\includegraphics{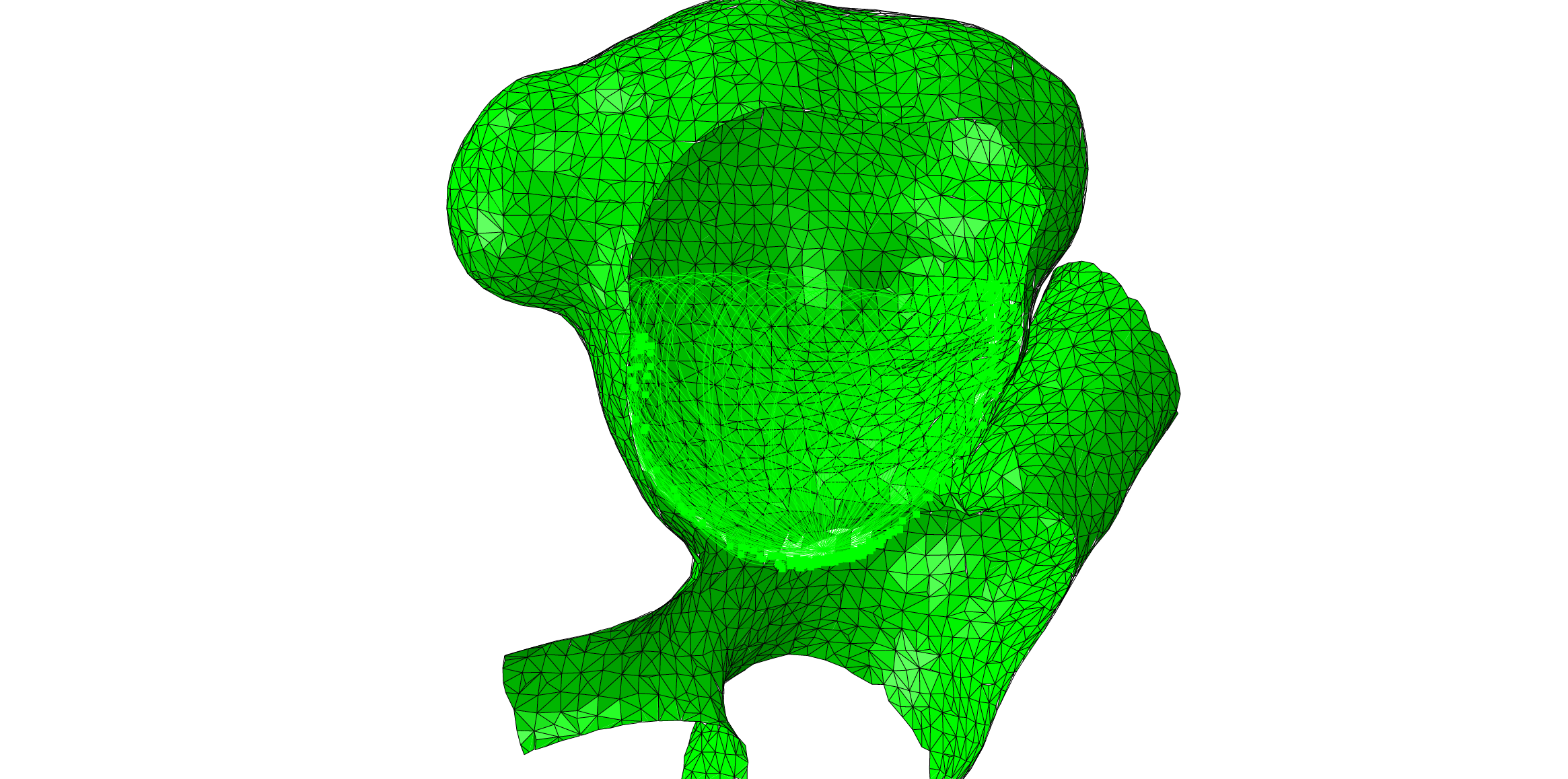}}
\caption{Shell mesh for Case II ($8,875$ elements)}
\label{fig:case_2}
\end{subfigure}
\hfill
\begin{subfigure}[b]{0.325\textwidth}
\centering
\adjustbox{width=1.4\textwidth, center}{\includegraphics{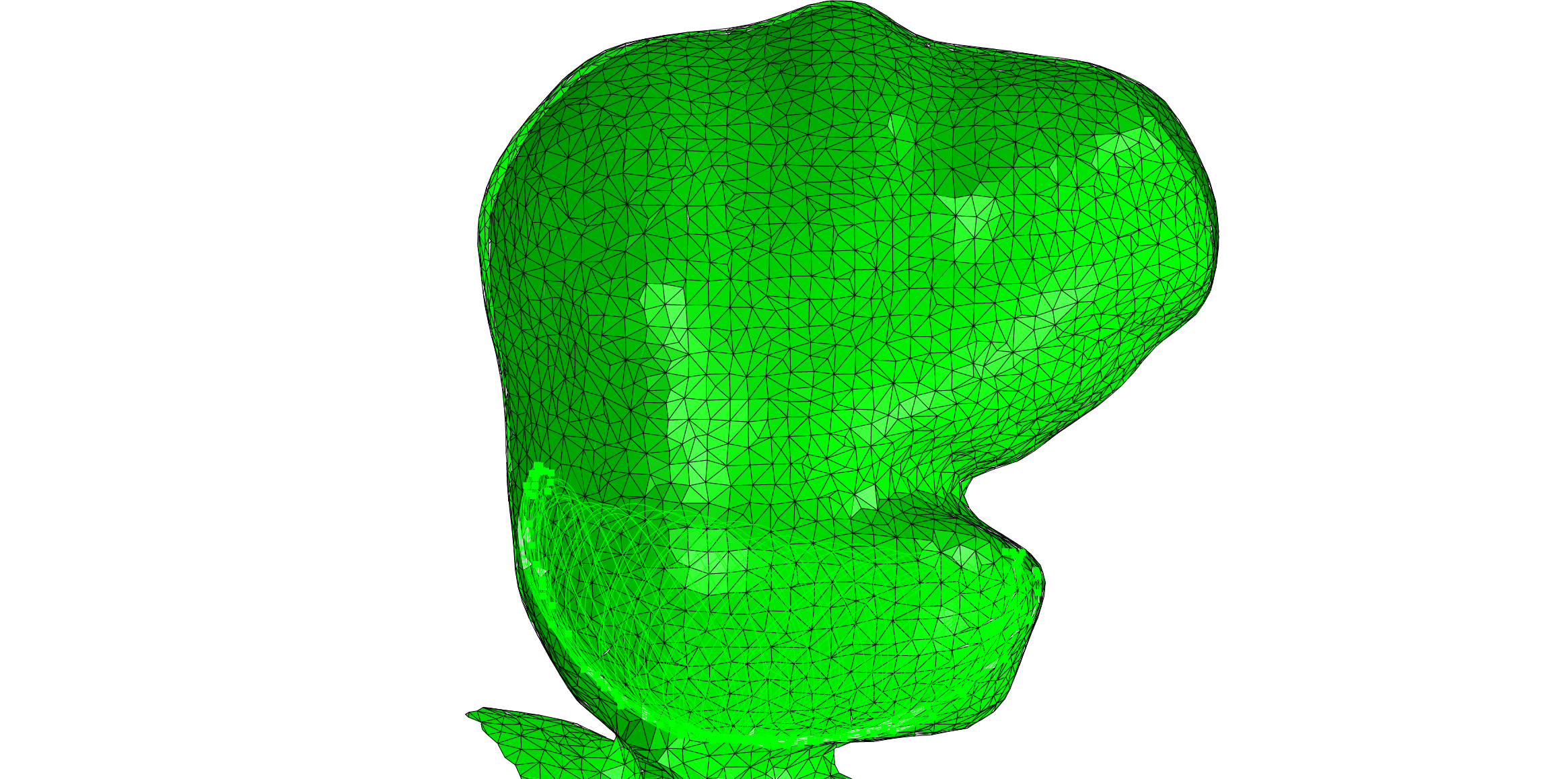}}
\caption{Shell mesh for Case III ($9,342$ elements)}
\label{fig:case_3}
\end{subfigure}
\caption{Isolated and discretized patient-specific aneurysm domains utilized as deformable shell environments in the explicit deployment simulations. The models are trimmed to retain only the local sac and short parent-vessel segments required to enforce structural boundary conditions.}
\label{fig:meshed_aneurysms}
\end{figure*}

The pre-treatment 3D-RA data are used to build the structural deployment models. The post-treatment CTA data are used only to guide the qualitative fast-placement reconstructions described in Section~\ref{subsec:fast_placement}. Vascular structures are segmented in MeVisLab v3.4.1 (MeVis Medical Solutions AG, Bremen, Germany). The resulting lumen surfaces are post-processed in Blender v4.2 (Blender Foundation, Amsterdam, The Netherlands): imaging artifacts are removed, surfaces are smoothed, elongated outlets are trimmed, and the final surfaces are re-triangulated. For the deployment analyses, the lumen surfaces are converted into deformable patient-specific shell domains (Figure~\ref{fig:segmented_aneurysms}), which are subsequently isolated and discretized into local computational domains (Figure~\ref{fig:meshed_aneurysms}).

\subsection{CNS device representation}
\label{subsec:device_model}

\begin{figure}[htbp!]
\centering

\begin{subfigure}[b]{\columnwidth}
\centering
\includegraphics[width=\linewidth]{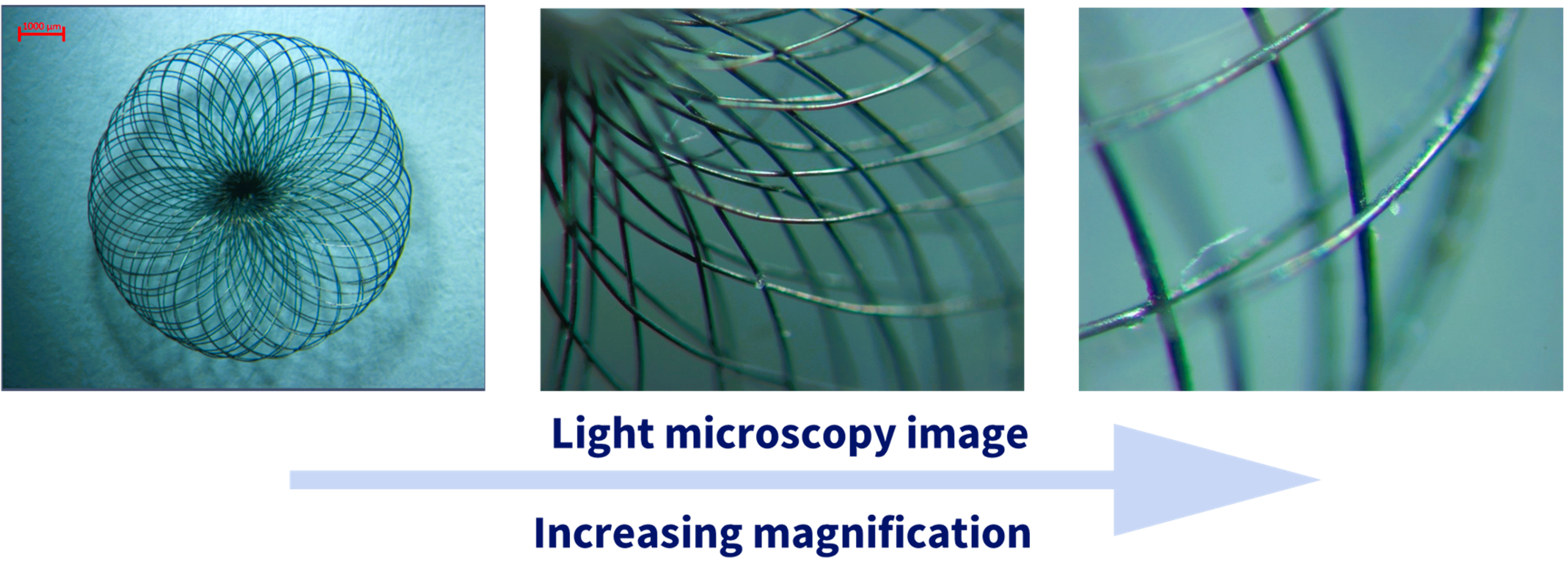}
\caption{Progressive magnification reveals the local interwoven wire architecture and crossover pattern; the red scale bar (top left) denotes $1000~\mu\mathrm{m}$.}
\label{fig:cns_microscopy}
\end{subfigure}

\vspace{2mm}

\begin{subfigure}[b]{\columnwidth}
\centering
\includegraphics[width=\linewidth]{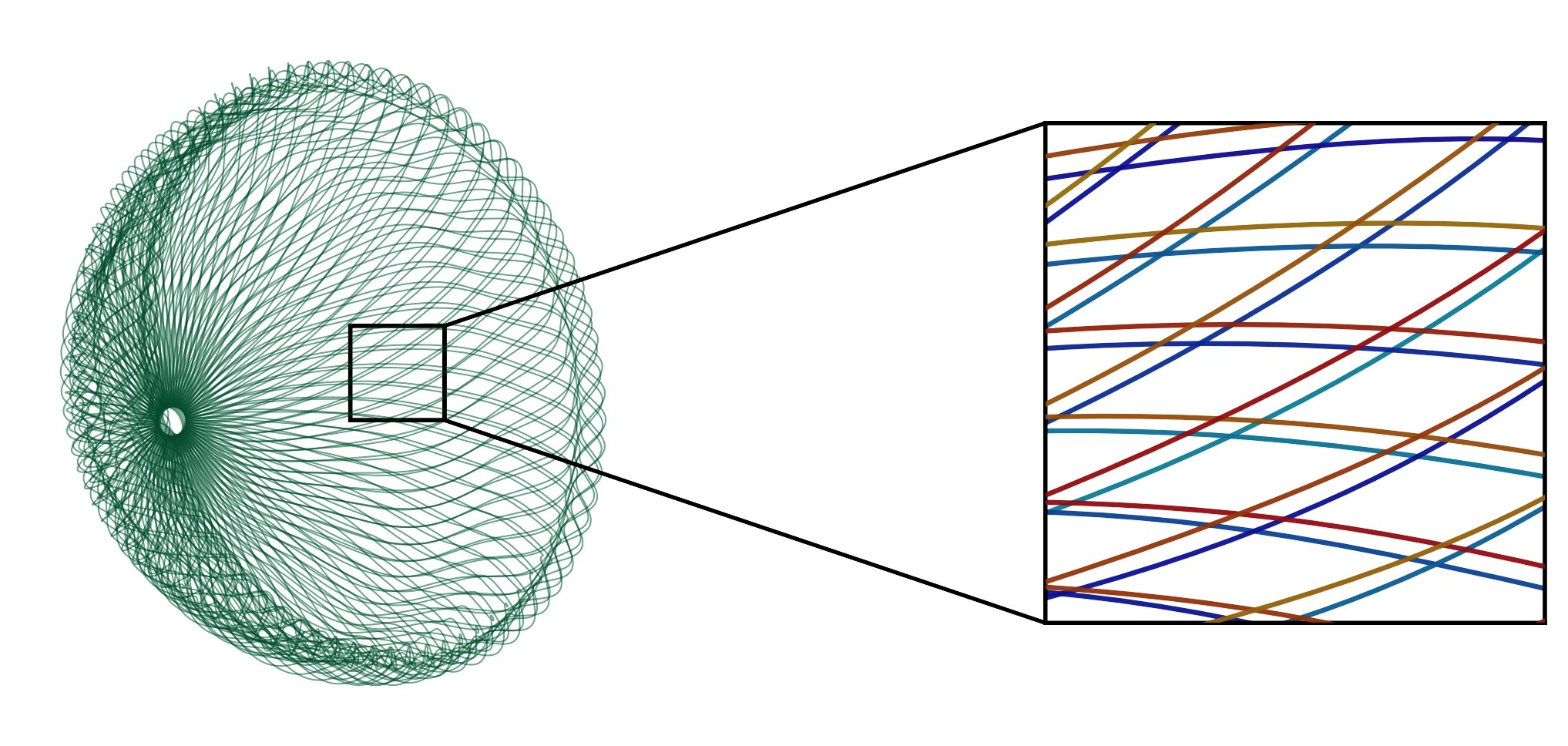}
\caption{Generated wire-resolved reference braid, with an enlarged interior
region resolving individual wire trajectories and crossings.}
\label{fig:contour_braid_geometry}
\end{subfigure}

\vspace{2mm}

\begin{subfigure}[b]{0.48\columnwidth}
\centering
\adjustbox{trim=100pt 0pt 100pt 0pt, clip, width=0.92\linewidth}{%
\includegraphics{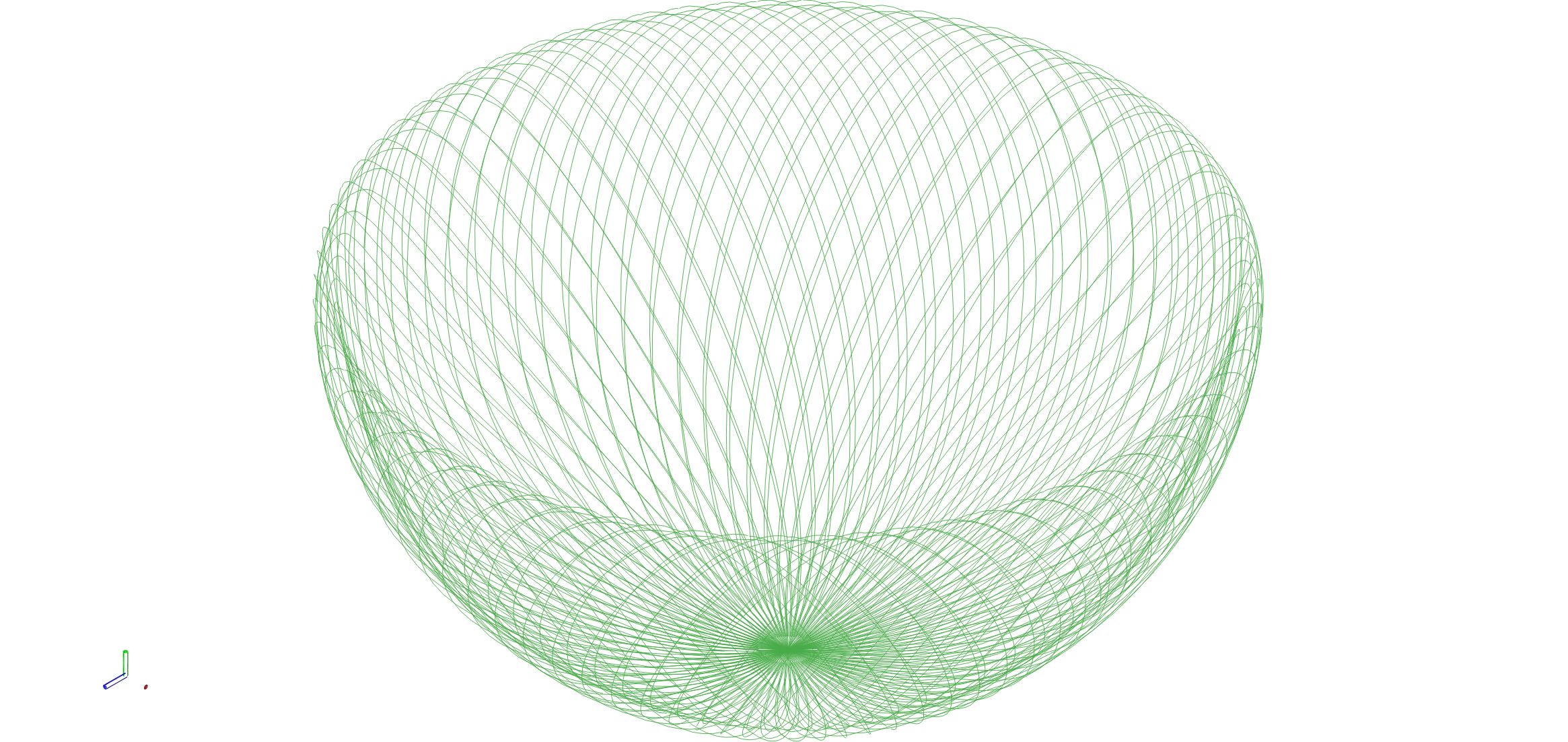}
}
\caption{Stress-free undeployed braid.}
\label{fig:cns_undeployed}
\end{subfigure}
\hfill
\begin{subfigure}[b]{0.48\columnwidth}
\centering
\adjustbox{trim=275pt 0pt 275pt 0pt, clip, width=0.62\linewidth}{%
\includegraphics{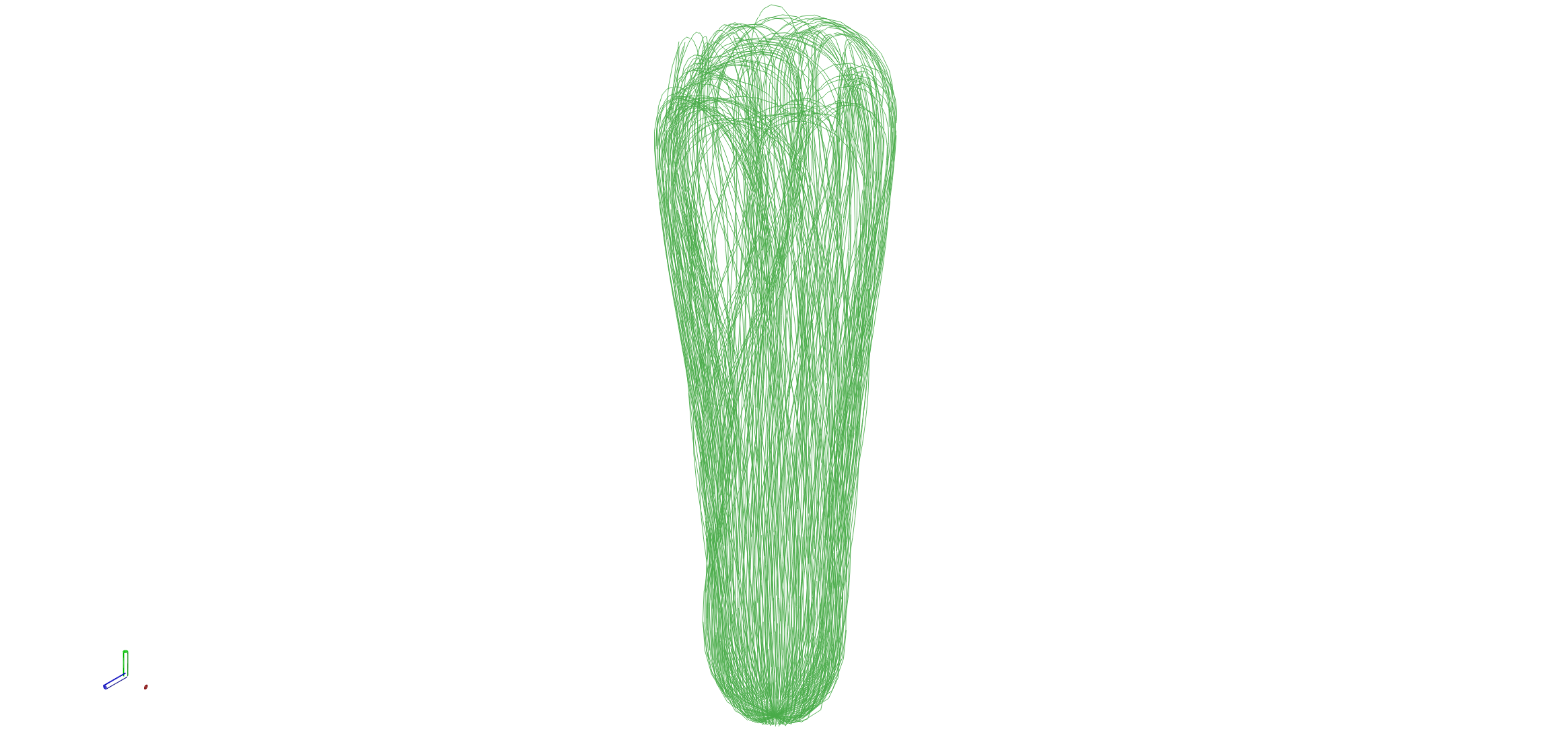}
}
\caption{Crimped configuration.}
\label{fig:cns_crimped}
\end{subfigure}

\caption{Overview of the CNS device representation. (a) Light microscopy
shows the local interwoven architecture of the physical braid. (b) The
corresponding wire-resolved computational reference geometry. (c,d) The
stress-free undeployed and crimped configurations used before staged release.
The final implanted geometry is not prescribed from these reference states; it
is obtained from the contact-resolved deployment analysis.}
\label{fig:cns_device_model}
\end{figure}

The complete device-model construction is summarized in Figure~\ref{fig:cns_device_model}. Light microscopy illustrates the local interwoven architecture of the physical device, whereas the computational reference geometry is generated from the parametric construction described in Section~\ref{subsec:braid_geometry} and discretized as individual beam continua using BeamMe \cite{BeamMe}.

The model preserves continuous wire paths and local crossing topology without imposing any kinematic constraints at the wire crossovers. The relative motion at crossings is instead governed by the subsequent contact formulation. This is essential because braid compaction, local rearrangement, wall apposition, anchoring, and neck seating depend on contact-mediated slip between neighboring wires \cite{shiozaki2021computational}. The undeployed braid defines the stress-free reference state. The crimped configuration is obtained from a preceding mechanical crimping and catheter-positioning simulation rather than by geometrically scaling the undeployed device. The resulting deformed configuration, together with the corresponding history-dependent superelastic material state, is transferred to the subsequent deployment analysis.

Consequently, deployment begins from a mechanically preconditioned state that contains the deformation history accumulated during crimping. The final implanted configuration is, therefore, neither prescribed nor generated from a stress-free crimped geometry; it emerges from recovery of the preconditioned braid together with evolving wire-wire, device-wall, and device-catheter contact.

\subsection{Imaging-guided geometric reconstruction}
\label{subsec:fast_placement}

To provide a standard baseline representing common imaging-guided virtual placement workflows, we generate qualitative geometric reconstructions in Blender. The nominal device geometry is manually positioned inside the pre-treatment aneurysm dome using the post-treatment CTA and the radiopaque CNS marker as anatomical guidance. We then apply the Blender Lattice Modifier to deform the device mesh until it visually conforms to the observed device pose and the segmented aneurysm-wall boundary.

This approach mirrors standard image-based reconstruction and fast virtual placement workflows frequently employed in retrospective hemodynamic assessments \cite{bernovskis2026hemodynamic,giessler2026clinical}. Rather than serving as a mechanically validated prediction, this geometric reconstruction is utilized to demonstrate the specific physical details, such as wire-wire self-contact, wire-wall impenetrability, tangential slip resistance, local contact tractions, and the resulting wall-supported state, that are captured exclusively through a contact-resolved finite element framework.

\section{Deployment formulation}
\label{sec:deployment_formulation}
This section defines the computational model used to simulate CNS deployment within the patient-specific aneurysm geometries. Stress-free reference centerlines are first generated for the Nitinol wires to construct the undeployed dual-layer braid. The braid is subsequently crimped, positioned, and released against the deformable aneurysm wall. The resulting configuration emerges from superelastic recovery, large deformation, wire-wire and wire-wall contact, frictional sliding, and wall support.

The parametric relations define only the reference geometry of the individual wires. The parametric description determines each wire's global trajectory and circumferential phase through polar and azimuthal relations, while radial and transverse perturbations generate the characteristic woven architecture. The following sections then describe the coupled structural model comprising the braid, deformable wall, superelastic material response, and frictional contact interactions.

\subsection{Braid parametrization}
\label{subsec:braid_geometry}

The computational CNS reference geometry is represented as a dual-layer, wire-resolved braid comprising $n_{\mathrm{w}}=72$ Nitinol wires per layer, consistent with the architecture adopted from inspection of the physical device. The two layers are assigned opposite azimuthal handedness so that their wire trajectories intersect and form the local braided architecture. For wire index $i\in\{0,\ldots,n_{\mathrm{w}}-1\}$ in layer $\ell\in\{0,1\}$, let $\sigma_\ell=+1$ for the first layer and $\sigma_\ell=-1$ for the second layer. The circumferential phase assigned to each wire is
\begin{equation}
\phi_{i\ell}
=
\frac{2\pi i}{n_{\mathrm{w}}}
+
\ell\,\Delta\phi ,
\label{eq:layer_phase}
\end{equation}
where $\Delta\phi$ denotes the relative phase shift between the two layers; the present model uses $\Delta\phi=1^\circ$. The wire paths are described by the normalized coordinate along the wire centerline $f\in[0,1]$. In the numerical implementation, the interval $f\in[0.01,0.99]$ is sampled to avoid evaluating the spherical parametrization exactly at the common pole.

The polar angle and the layer-dependent azimuthal angle are defined as
\begin{equation}
\vartheta(f)
=
\frac{\pi}{2}\sin(\pi f),
\qquad
\varphi_{i\ell}(f)
=
\sigma_\ell\frac{\pi}{2}f+\phi_{i\ell}.
\label{eq:angles}
\end{equation}
Thus, the two layers follow opposite azimuthal trajectories. The radial coordinate is modulated by a tapered undulation,
\begin{equation}
r_\ell(f)
=
R+
\sigma_\ell
\sin(\pi f)\,A_r\,
\sin\!\left(2\pi W_r f^{p_r}\right),
\label{eq:radiallaw}
\end{equation}
where $R$ is the nominal device radius, while $A_r$, $W_r$, and $p_r$ control the amplitude, number, and longitudinal distribution of the radial undulations, respectively.

The resulting base centerline of wire $(i,\ell)$ is
\begin{equation}
\vect{c}_{i\ell}(f)
=
r_\ell(f)
\begin{pmatrix}
\sin\vartheta(f)\cos\varphi_{i\ell}(f)\\
-\cos\vartheta(f)\\
\sin\vartheta(f)\sin\varphi_{i\ell}(f)
\end{pmatrix}.
\label{eq:basecurve}
\end{equation}
Its unit tangent is
\begin{equation}
\vect{t}_{i\ell}(f)
=
\frac{\dd\vect{c}_{i\ell}/\dd f}
{\norm{\dd\vect{c}_{i\ell}/\dd f}}.
\label{eq:tangent}
\end{equation}

A local transverse direction is constructed from the radial unit vector
\begin{equation}
\vect{e}_{r,i\ell}(f)
=
\frac{\vect{c}_{i\ell}(f)}
{\norm{\vect{c}_{i\ell}(f)}}.
\label{eq:radial_direction}
\end{equation}
Because $\vect{e}_{r,i\ell}$ is not generally orthogonal to the wire tangent, it is projected onto the normal plane of the base centerline:
\begin{equation}
\vect{n}_{i\ell}(f)
=
\frac{
\vect{e}_{r,i\ell}(f)
-
\left[
\vect{e}_{r,i\ell}(f)\cdot\vect{t}_{i\ell}(f)
\right]\vect{t}_{i\ell}(f)
}{
\norm{
\vect{e}_{r,i\ell}(f)
-
\left[
\vect{e}_{r,i\ell}(f)\cdot\vect{t}_{i\ell}(f)
\right]\vect{t}_{i\ell}(f)
}
}.
\label{eq:weave_normal}
\end{equation}
The vector $\vect{n}_{i\ell}$ is therefore transverse to the base-wire centerline while remaining as close as possible to the local radial direction.

The final undeployed centerline is obtained by imposing a tapered, oscillatory perturbation in this transverse direction:
\begin{equation}
\vect{x}_{i\ell}(f)
=
\vect{c}_{i\ell}(f)
+
\sigma_\ell
\left[\sin(\pi f)\right]^{p_w}A_w
\sin\!\left(2\pi W_w f+\phi_{i\ell}\right)
\vect{n}_{i\ell}(f).
\label{eq:centerline}
\end{equation}
Here, $A_w$, $W_w$, and $p_w$ control the amplitude, number, and tapering of the local weave perturbation. Equation~\eqref{eq:centerline} defines the stress-free reference braid geometry before crimping and release.

\subsection{Structural formulation and discretization}
\label{subsec:structural_discretization}

Each wire is modeled as a geometrically exact Simo-Reissner beam \cite{simo1985beam,simo1986rod,simo1988dynamics,meier2019geometrically}. Let $s\in[0,L]$ denote arc length along the beam centerline, $\vect{r}(s)$ the beam centerline, and $\Lambda(s)\in\mathrm{SO}(3)$ the cross-section triad. The generalized strain measures are
\begin{equation}
\vect{\Gamma}
=
\Lambda^{\mathsf T}\vect{r}_{,s}-\vect{e}_1,
\qquad
\vect{K}
=
\axl\!\left(\Lambda^{\mathsf T}\Lambda_{,s}\right)-\vect{\kappa}_0,
\label{eq:beamstrains}
\end{equation}
Here, $\axl(\cdot)$ denotes the axial-vector operator that maps a skew-symmetric tensor to its associated vector. $\vect{\Gamma}$ contains axial and shear strains, while $\vect{K}$ contains bending and torsional strains. The vector $\vect{\kappa}_0$ denotes the reference curvature of the generated undeployed braid. The undeployed braid geometry is taken as stress-free before crimping; hence, subtracting $\vect{\kappa}_0$ prevents the initially curved wires from carrying artificial bending energy. The internal virtual work of each beam is written in resultant form as
\begin{equation}
\mathcal{G}_{\mathrm{beam}}
=
\int_0^L
\left(
\delta\vect{\Gamma}\cdot\vect{n}
+
\delta\vect{K}\cdot\vect{m}
\right)\dd s,
\label{eq:beamweak}
\end{equation}
where $\vect{n}$ and $\vect{m}$ are the cross-sectional force and moment resultants associated with the axial/shear and bending/torsional strain measures, respectively. Unlike a linear-elastic beam model, these resultants are not related to $\vect{\Gamma}$ and $\vect{K}$ through constant constitutive matrices. They are obtained from the history-dependent superelastic constitutive response of the Nitinol wire section and therefore depend on the current strain measures and internal transformation state. This formulation resolves the large rotations, bending, torsion, and local rearrangements expected during CNS release.

Let \(\mathbf q\) denote the generalized configuration of the coupled beam-wall system, and let \(\mathcal V_0\) denote the space of kinematically admissible virtual variations that vanish on prescribed displacement and rotation boundaries. Collecting the beam, wall, contact, inertial, damping, and external contributions, the coupled deployment problem is stated in weak form as
\begin{equation}
\mathcal{G}(\delta\vect{q},\vect{q})
=
\mathcal{G}_{\mathrm{dyn}}+
\mathcal{G}_{\mathrm{damp}}+
\mathcal{G}_{\mathrm{beam}}+
\mathcal{G}_{\mathrm{wall}}+
\mathcal{G}_{\mathrm{cont}}-
\mathcal{G}_{\mathrm{ext}}
=0
\quad \forall\,\delta\vect{q}\in\mathcal{V}_0.
\label{eq:weakform}
\end{equation}
Here, $\mathcal{G}_{\mathrm{dyn}}$, $\mathcal{G}_{\mathrm{damp}}$, $\mathcal{G}_{\mathrm{beam}}$, $\mathcal{G}_{\mathrm{wall}}$, $\mathcal{G}_{\mathrm{cont}}$, and $\mathcal{G}_{\mathrm{ext}}$ denote the virtual-work contributions of inertia, numerical damping, beam deformation, wall deformation, contact interactions, and external loading, respectively.

For the finite-element approximation, the beam and shell continua introduced above are spatially discretized as follows. The $144$ CNS wires are discretized using $10{,}512$ three-node quadratic beam elements (Abaqus element type: B32), corresponding to $73$ elements per wire. The aneurysm-wall meshes comprised $1{,}897$ nodes and $3{,}717$ three-node triangular shell elements for Case~I, $4{,}465$ nodes and $8{,}875$ elements for Case~II, and $4{,}735$ nodes and $9{,}342$ elements for Case~III (Abaqus element type: S3), as visualized in Figure~\ref{fig:meshed_aneurysms}. Thus, the device and aneurysm wall form a mixed-dimensional structural system \cite{datz2025patient,datz2026influence}. After spatial discretization, the semi-discrete equations read
\begin{equation}
\vect{r}_{\mathrm{dyn}}(\vect{q},\dot{\vect{q}},\ddot{\vect{q}})
+\vect{f}_{\mathrm{damp}}(\vect{q},\dot{\vect{q}})
+\vect{f}_{\mathrm{int}}^{\mathrm{beam}}(\vect{q})
+\vect{f}_{\mathrm{int}}^{\mathrm{wall}}(\vect{q})
+\vect{f}_{\mathrm{cont}}(\vect{q},\dot{\vect{q}})
=
\vect{f}_{\mathrm{ext}}(t),
\label{eq:semidiscrete}
\end{equation}
where the discretized vector $\vect{q}$ collects the translational and rotational device degrees of freedom together with the wall degrees of freedom, and an overdot denotes differentiation with respect to time. $\vect{r}_{\mathrm{dyn}}$ denotes the inertial force vector, including mass-acceleration and configuration-dependent contributions associated with finite rotations; $\vect{f}_{\mathrm{damp}}$ denotes numerical damping; and $\vect{f}_{\mathrm{cont}}$ collects the wire-wire, device-wall, and device-catheter contact forces. The deformable aneurysm wall provides the anatomical confinement that arrests device expansion during seating. All deployment analyses are performed in Abaqus/Explicit using geometrically nonlinear beam elements and three-node triangular shell elements with large-deformation contact.

\subsection{Material models}
\label{subsec:material_models}

The aneurysm wall is modeled as a hyperelastic shell using a reduced-polynomial Yeoh strain-energy density,
\begin{equation}
W(\bar I_1,J)
=
C_{10}(\bar I_1-3)+C_{20}(\bar I_1-3)^2+C_{30}(\bar I_1-3)^3+
\frac{1}{D_1}(J-1)^2,
\label{eq:yeoh}
\end{equation}
where $\bar I_1$ is the first deviatoric invariant and $J=\det(\vect{F})$ is the determinant of the deformation gradient, i.e., the ratio of the change of volume between material and current configurations. The wall contribution is
\begin{equation}
\mathcal{G}_{\mathrm{wall}}
=
\int_{\Omega_{\mathrm{wall}}}
\delta\vect{E}:\vect{S}\,\dd\Omega,
\label{eq:wallweak}
\end{equation}
where $\delta\vect{E}$ is the virtual shell strain and $\vect{S}$ is the corresponding second Piola-Kirchhoff stress measure. For the reduced-polynomial potential, the initial shear and bulk moduli obtained by linearization about the undeformed configuration are
\begin{equation}
G_0=2C_{10},
\qquad
K_0=\frac{2}{D_1},
\end{equation}
\begin{equation}
E_0=\frac{9K_0G_0}{3K_0+G_0},
\qquad
\nu_0=\frac{3K_0-2G_0}{2(3K_0+G_0)}.
\label{eq:initial_wall_moduli}
\end{equation}

\begin{table*}[t!]
\centering
\small
\caption{Key geometric, constitutive, and numerical parameters used in the deployment simulations.}
\label{tab:model_parameters}
\begin{tabularx}{\textwidth}{>{\raggedright\arraybackslash}p{3.2cm} >{\raggedright\arraybackslash}p{4.0cm} >{\centering\arraybackslash}p{2.6cm} >{\centering\arraybackslash}p{1.8cm} X}
\toprule
Group & Quantity & Symbol / value & Units & Role in the model \\
\midrule
Device geometry & Nominal CNS diameter & $d_{\mathrm{CNS}}=7.0$ & mm & Undeployed device size \\
Device geometry & Computational wire architecture & $2\times72=144$ wires & - & Dual-layer representation based on physical-device inspection \\
Device geometry & Reference radius & $R=3.5$ & mm & Nominal radius in the braid parametrization \\
Device geometry & Inter-layer phase shift & $\Delta\phi=1.0$ & deg & Relative circumferential phase between braid layers \\
Device geometry & Radial undulation & $A_r=0.12$, $W_r=7$, $p_r=0.1$ & mm, -, - & Radial modulation in Eq.~\eqref{eq:radiallaw} \\
Device geometry & Weave perturbation & $A_w=0.12$, $W_w=7$, $p_w=0.2$ & mm, -, - & Local weave modulation in Eq.~\eqref{eq:centerline} \\
Device geometry & Wire diameter & $d_w=0.020$ & mm & Circular Nitinol wire cross-section \\
Wall & Shell thickness & $t_{\mathrm{wall}}=0.50$ & mm & Deformable confinement \\
Wall & Yeoh coefficients & $C_{10}=0.80$, $C_{20}=2.20$, $C_{30}=6.00$ & MPa & Homogeneous surrogate wall; $C_{10}$ sets $G_0=1.60$~MPa, while $C_{20}$ and $C_{30}$ control finite-strain stiffening \\
Wall & Volumetric parameter & $D_1=0.04$ & MPa$^{-1}$ & $K_0=50.0$~MPa and $\nu_0\approx0.484$ \\
Numerics & Rayleigh damping & $\alpha=10$, $\beta=10^{-4}$ & s$^{-1}$, s & Attenuation of oscillatory response during explicit deployment; not a measured material property \\
Numerics & Explicit step durations & $1,\ 4,\ 20,\ 35$ & s & Numerical durations of the four sequential deployment stages \\
Numerics & Fixed mass scaling & $\Delta t_{\mathrm{target}}=10^{-4}$ & s & Applied below the minimum stable increment in each explicit step \\
Nitinol & Austenite modulus and Poisson ratio & $E_A=5.0\times10^4$, $\nu_A=0.33$ & MPa, - & Superelastic wire response \\
Nitinol & Martensite modulus and Poisson ratio & $E_M=2.8\times10^4$, $\nu_M=0.33$ & MPa, - & Transformed phase response \\
Nitinol & Transformation strain & $\varepsilon^{\mathrm{tr}}=0.06$ & - & Recoverable transformation strain \\
Nitinol & Loading transformation stresses & $\sigma_s^L=400$, $\sigma_f^L=450$ & MPa & Forward transformation interval \\
Nitinol & Unloading transformation stresses & $\sigma_s^U=250$, $\sigma_f^U=200$ & MPa & Reverse transformation interval \\
Nitinol & Compression-side onset & $\sigma_{s,c}^{L}=500$ & MPa & Compression transformation threshold \\
\bottomrule
\end{tabularx}
\end{table*}

Accordingly, the selected Yeoh coefficients define the initial elastic response and the progressive distortional stiffening at finite strain. The complete parameter set and the corresponding initial elastic moduli are listed in Table~\ref{tab:model_parameters}. These coefficients are not identified from the tissue of the three patients. They define a homogeneous surrogate wall selected to retain deformable confinement while avoiding excessive wall distortion during the highly nonlinear contact-release analysis. The parameterization is, therefore, a pragmatic literature-informed surrogate rather than a patient-specific constitutive fit. Experimental studies demonstrate substantial inter-patient and spatial variability in intracranial aneurysm-wall properties and report materially different Yeoh coefficients \cite{brunel2018rupture,oliveira2022numerical,parshin2019optimal}. Consequently, the present wall model is used to determine the wall-supported deployment configuration and is not intended to provide patient-specific predictions of aneurysm-wall stress or failure.

The Nitinol wires are modeled with an Auricchio-type superelastic law \cite{auricchio1997shape}. At the beam level, the constitutive response is evaluated over the circular wire cross-section to obtain the cross-sectional force and moment resultants, $\vect{n}$ and $\vect{m}$, as introduced in Equation~\eqref{eq:beamweak}. Thus, axial loading, bending, and the crimping-release cycle follow the same history-dependent superelastic material response. The adopted parameter set is a pragmatic literature-informed surrogate selected to produce a stable superelastic crimping-release response. It is not a device-specific thermomechanical calibration against CNS bench-test data \cite{elsisy2020materials}. Table~\ref {tab:model_parameters} lists the values used in the present simulations.

\subsection{Contact formulation}
\label{subsec:contact_formulation}

Contact is enforced at the wire-wire, device-wall, and device-catheter interfaces. For a generic contact pair, let $g_n$ be the signed normal gap, positive in separation and negative in penetration. The scalar normal contact traction is
\begin{equation}
t_n=\varepsilon_n\langle -g_n\rangle,
\qquad
\langle x\rangle=\frac{1}{2}(x+|x|),
\label{eq:normalcontact}
\end{equation}
with traction vector $\vect{t}_n=t_n\vect{n}$. Normal contact is enforced using the default Abaqus/Explicit linear penalty formulation, where $\varepsilon_n$ denotes the corresponding penalty stiffness. The normal contact potential is
\begin{equation}
\Pi_{\mathrm{cont}}^{n}
=
\int_{\Gamma_c}\frac{1}{2}\varepsilon_n\langle -g_n\rangle^2\,\dd\Gamma .
\label{eq:normalpotential}
\end{equation}

Tangential contact follows a regularized Coulomb law. Let $\vect{g}_t$ denote the accumulated tangential slip measure. In stick, the trial traction is
\begin{equation}
\vect{t}_t^{\mathrm{trial}}=-\varepsilon_t\vect{g}_t,
\label{eq:trialstick}
\end{equation}
subject to
\begin{equation}
\Phi(\vect{t}_t,t_n)=\norm{\vect{t}_t}-\mu t_n\le 0.
\label{eq:yield}
\end{equation}
When $\Phi>0$, sliding is described by
\begin{equation}
\vect{t}_t=-\mu t_n\frac{\dot{\vect{g}}_t}{\norm{\dot{\vect{g}}_t}}.
\label{eq:slidinglaw}
\end{equation}
The corresponding weak contact contribution is
\begin{equation}
\mathcal{G}_{\mathrm{cont}}
=
\int_{\Gamma_c}\delta g_n\,t_n\,\dd\Gamma+
\int_{\Gamma_c}\delta\vect{g}_t\cdot\vect{t}_t\,\dd\Gamma .
\label{eq:contactweak}
\end{equation}
Thus, normal impenetrability, tangential stick, and sliding friction enter the same deployment solution.

\begin{figure}[htpb!]
\centering
\includegraphics[width=0.6\linewidth]{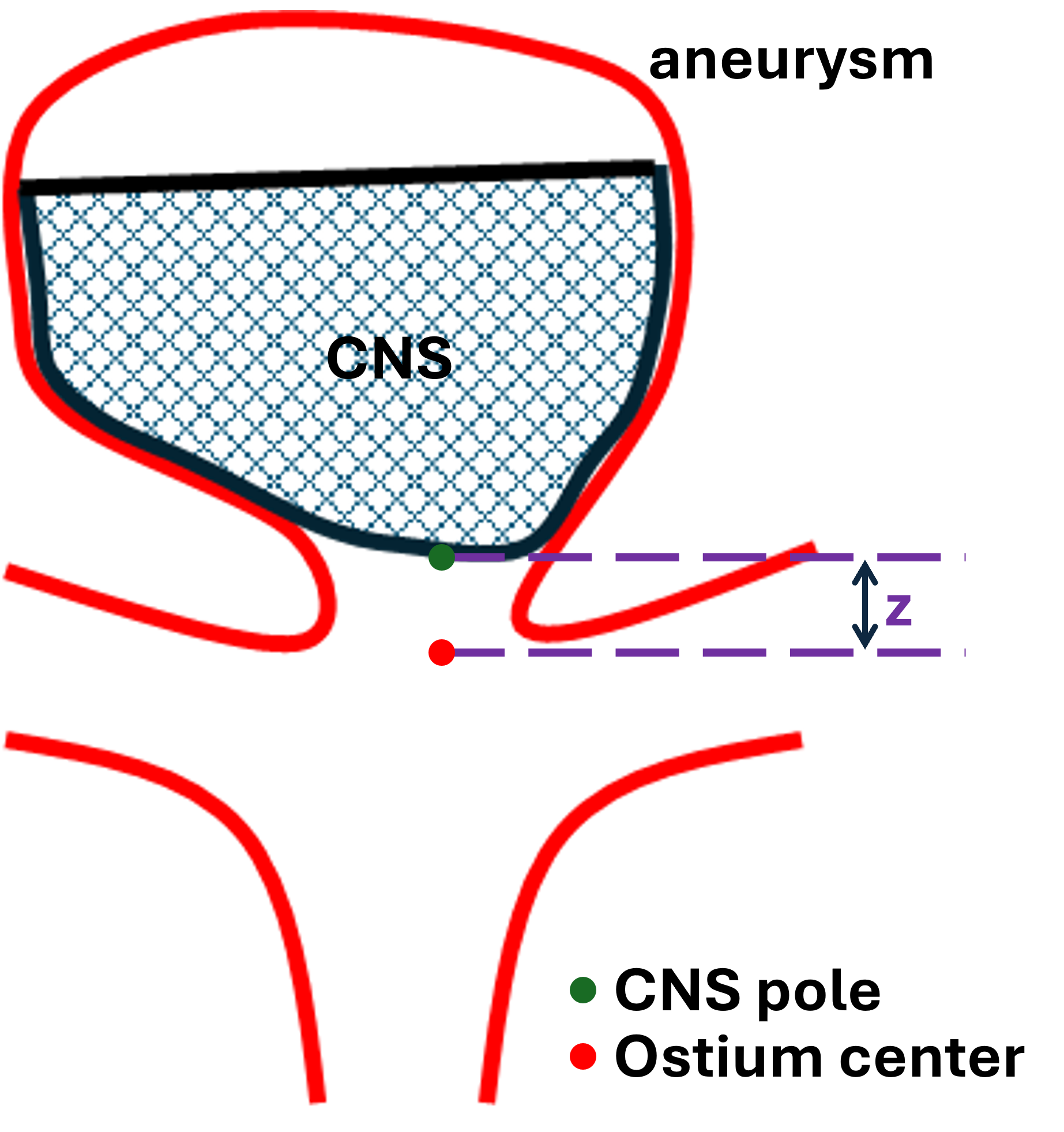}
\caption{Definition of the release-depth parameter $z$. The signed axial distance between the CNS pole and the aneurysm neck plane is measured along the local release direction. Positive and negative values describe neck-adjacent and deeper sac-side initial release positions, respectively.}
\label{fig:release_height_schematic}
\end{figure}

\subsection{Damping, contact conditions, and staged release}
\label{subsec:contact_release}

Rayleigh damping attenuates the oscillatory response during the explicit deployment analysis and promotes a slowly evolving release response. It is not intended to represent physiological dissipation in the device-wall system. The mass-proportional coefficient, $\alpha=10~\mathrm{s}^{-1}$, primarily attenuates low-frequency oscillatory motion, whereas the stiffness-proportional coefficient, $\beta=10^{-4}~\mathrm{s}$, attenuates higher-frequency response. Neither coefficient is interpreted as a measured property of the aneurysm wall or the Nitinol device.

Friction plays two mechanically distinct roles during CNS deployment. At wire-wire crossovers, low tangential resistance permits the individual wires to slide and rearrange as the braid expands and changes shape. At the device-wall interface, tangential resistance opposes sliding of the expanded device along the aneurysm wall and, therefore, contributes to its wall-supported seating. These mechanisms cannot be represented by a purely geometric placement, in which the implanted configuration is prescribed without resolving the forces and relative motion that develop during deployment. The present study treats wire-wire and device-wall friction separately: wire-wire friction is kept fixed at a low value to permit braid rearrangement, whereas the device-wall tangential condition is varied systematically to determine how strongly wall sliding and anchoring influence the mechanically realized deployed state.

Wire-wire contact is assigned a constant low friction coefficient,
\begin{equation}
\mu_{\mathrm{ww}}=0.01,
\end{equation}
throughout all simulations. This choice permits relative slip and local rearrangement at braid crossovers during crimping, expansion, and seating. Here, seating denotes the post-release process by which the device establishes wall-supported contact and approaches a late-time configuration (see Figure~\ref{fig:release_height_schematic}).

\begin{table*}[htbp!]
\centering
\small
\caption{Contact conditions used to span different regimes of device-wall tangential interaction, from frictionless sliding to the limiting no-slip condition. Wire-wire friction is fixed at $\mu_{\mathrm{ww}}=0.01$ in all cases. The prescribed device-wall conditions are used to assess the mechanical influence of tangential slip resistance and are not calibrated physiological properties of the CNS-aneurysm-wall interface.}
\label{tab:friction_cases}
\begin{tabularx}{\textwidth}{
>{\raggedright\arraybackslash}p{1.1cm}
>{\centering\arraybackslash}p{1.1cm}
>{\centering\arraybackslash}p{1.1cm}
>{\centering\arraybackslash}p{5.8cm}
Y}
\toprule
Case & $\mu_{\mathrm{ww}}$ & $\mu_{\mathrm{dw}}$ & Damping treatment & Mechanical interpretation \\
\midrule
\texttt{p0}   & 0.01 & 0     & baseline Rayleigh & Frictionless device-wall sliding lower bound \\
\texttt{p1}   & 0.01 & 1     & baseline Rayleigh & Finite device-wall resistance case \\
\texttt{p1d1} & 0.01 & 1     & Rayleigh + contact damping coefficient $1.0$ & Same device-wall friction as \texttt{p1} with additional numerical contact damping \\
\texttt{p3d1} & 0.01 & 3     & Rayleigh + contact damping coefficient $1.0$ & Strong device-wall tangential friction with additional numerical contact damping \\
\texttt{p5d1} & 0.01 & 5     & Rayleigh + contact damping coefficient $1.0$ & High device-wall resistance with additional numerical contact damping \\
\texttt{r}    & 0.01 & rough & baseline Rayleigh & No-slip device-wall limiting condition \\
\bottomrule
\end{tabularx}
\end{table*}

In contrast, the device-wall friction coefficient, $\mu_{\mathrm{dw}}$, is treated as the parameter controlling tangential resistance between the expanding braid and the aneurysm wall. Because no validated \textit{in vivo} value is available for the blood-wetted CNS-aneurysm-wall interface, the objective is not to identify or prescribe a physiological coefficient of friction. Instead, the device-wall contact conditions are chosen to span mechanically distinct regimes from unrestricted tangential sliding to complete suppression of sliding. This allows the role of device-wall friction in the deployment solution to be isolated and quantified.

All simulations include the baseline Rayleigh damping defined above, with $\alpha=10~\mathrm{s}^{-1}$ and $\beta=10^{-4}~\mathrm{s}$. Cases carrying the suffix \texttt{d1} additionally employ the Abaqus/Explicit contact-damping option
\texttt{*Contact Damping, definition=DAMPING COEFFICIENT}
with a damping coefficient of $1.0$ at the device-wall interface. This contact-damping contribution is numerical and is distinct from the baseline Rayleigh damping. Thus, \texttt{p1} and \texttt{p1d1} have the same device-wall friction coefficient, $\mu_{\mathrm{dw}}=1$, and differ only through the additional contact damping. Their comparison therefore isolates the sensitivity of the computed deployment response to this numerical dissipation \cite{abaqus2024documentation}.

Table~\ref{tab:friction_cases} summarizes the investigated contact conditions. The suffix \texttt{d1} denotes simulations with additional contact damping. All simulations include the baseline Rayleigh damping defined above; cases without this suffix use the baseline damping setting. During initial settling and constrained expansion, a low device-wall coefficient of $\mu_{\mathrm{dw}}=0.01$ is used to permit expansion and conformation of the braid. During final seating, the device-wall tangential condition is varied from frictionless contact ($\mu_{\mathrm{dw}}=0$), through progressively stronger finite friction ($\mu_{\mathrm{dw}}=1,3,5$), to rough contact. In Abaqus, rough contact represents the limiting no-slip condition in which tangential sliding at the device-wall interface is suppressed. Consequently, these cases constitute a controlled sweep from freely sliding to increasingly constrained device-wall interaction.

For the release-depth sensitivity study, the initial pole position is varied over
\[
z\in\{-1.0,-0.8,-0.6,-0.4,-0.2,0,0.2,0.4,0.6,0.8,1.0\}~\mathrm{mm},
\]
where $z$ is the signed axial distance from the aneurysm neck plane defined in Figure~\ref{fig:release_height_schematic}. The device-wall interface is assigned rough contact for this sweep, thereby suppressing tangential sliding during final seating. All other model parameters are held fixed.

The investigated device-wall friction levels should be interpreted as prescribed contact regimes used to probe the mechanical role of tangential resistance, rather than as estimates of the physiological CNS-aneurysm-wall coefficient of friction. Frictional interaction in related neurovascular devices depends on device design, surface finish, tissue or phantom material, lubrication, contact pressure, curvature, and sliding velocity \cite{kelly2019comparison,shiozaki2021computational,otani2020modeling,nagargoje2025impact,elkhayyat2024friction}. In the absence of device-specific, blood-wetted interface calibration, the cases $\mu_{\mathrm{dw}}=3$, $\mu_{\mathrm{dw}}=5$, and rough contact are interpreted as progressively stronger wall-anchoring conditions, with rough contact providing the idealized no-slip bound, rather than as candidate coefficients for patient-specific prediction.

The deployment analysis starts from an imported preloaded configuration representing the preceding crimping and catheter-positioning procedure. After initial settling, the device is advanced and allowed to expand against the aneurysm wall while its pole remains temporarily constrained. In the final seating stage, the remaining pole constraint is removed, so that all device degrees of freedom are free except where device-wall contact, wire-wire contact, friction, and catheter contact act. The final configuration is, therefore, not prescribed, but emerges from superelastic recovery and the evolving contact interactions.

The four sequential Abaqus/Explicit steps are assigned numerical durations of $1$, $4$, $20$, and $35$, respectively. Fixed mass scaling is applied in each step to elements whose stable time increment falls below a target value of $\Delta t_{\mathrm{target}}=10^{-4}$. These settings are introduced for computational efficiency and numerical stabilization and are not interpreted as physical deployment times or material parameters.

\subsection{Deployment metrics}
\label{subsec:metrics}

The primary history variables are the device-wall contact area $A_{\mathrm{contact}}(t)$ and the monitored pole displacement $u_{\mathrm{tip}}(t)$. Final neck engagement is quantified along the neck rim $\Gamma_{\mathrm{neck}}$. Let $s$ denote arc length along the neck rim, $L_{\mathrm{neck}}$ its total length, and $\chi_c(s)$ an indicator equal to one where the rim is in contact with the device and zero otherwise. The rim coverage fraction is
\begin{equation}
\eta_{\mathrm{cov}}
=
\frac{1}{L_{\mathrm{neck}}}
\int_{\Gamma_{\mathrm{neck}}}\chi_c(s)\,\dd s.
\label{eq:coverage}
\end{equation}
The largest continuously uncovered rim segment is
\begin{equation}
g_{\max}
=
\max_{\mathcal{U}\subset\Gamma_{\mathrm{neck}}}
\int_{\mathcal{U}}\left[1-\chi_c(s)\right]\,\dd s,
\label{eq:gap}
\end{equation}
where $\mathcal{U}$ denotes a connected uncovered arc. For a compact scalar estimate of the engaged neck area, an equivalent circular neck area and its contacted fraction are calculated as
\begin{equation}
A_{\mathrm{neck,eq}}
=
\frac{L_{\mathrm{neck}}^2}{4\pi},
\qquad
A_{\mathrm{cov,eq}}
=
\eta_{\mathrm{cov}}A_{\mathrm{neck,eq}}.
\label{eq:areageom}
\end{equation}
Here, $A_{\mathrm{cov,eq}}$ is a coverage-equivalent area descriptor obtained from the neck-rim coverage fraction; it is not a measure of physical device-wall contact area. Thus, $\eta_{\mathrm{cov}}=1$ and $g_{\max}=0$ correspond to complete neck-rim coverage, whereas reduced $\eta_{\mathrm{cov}}$ and increased $g_{\max}$ indicate incomplete ostium engagement. These metrics distinguish neck coverage from generic device-wall contact within the aneurysm sac.

\section{Results}
\label{sec:results}

\begin{figure*}[htbp!]
\centering

\begin{subfigure}[t]{0.48\textwidth}
\centering
\includegraphics[width=\linewidth]{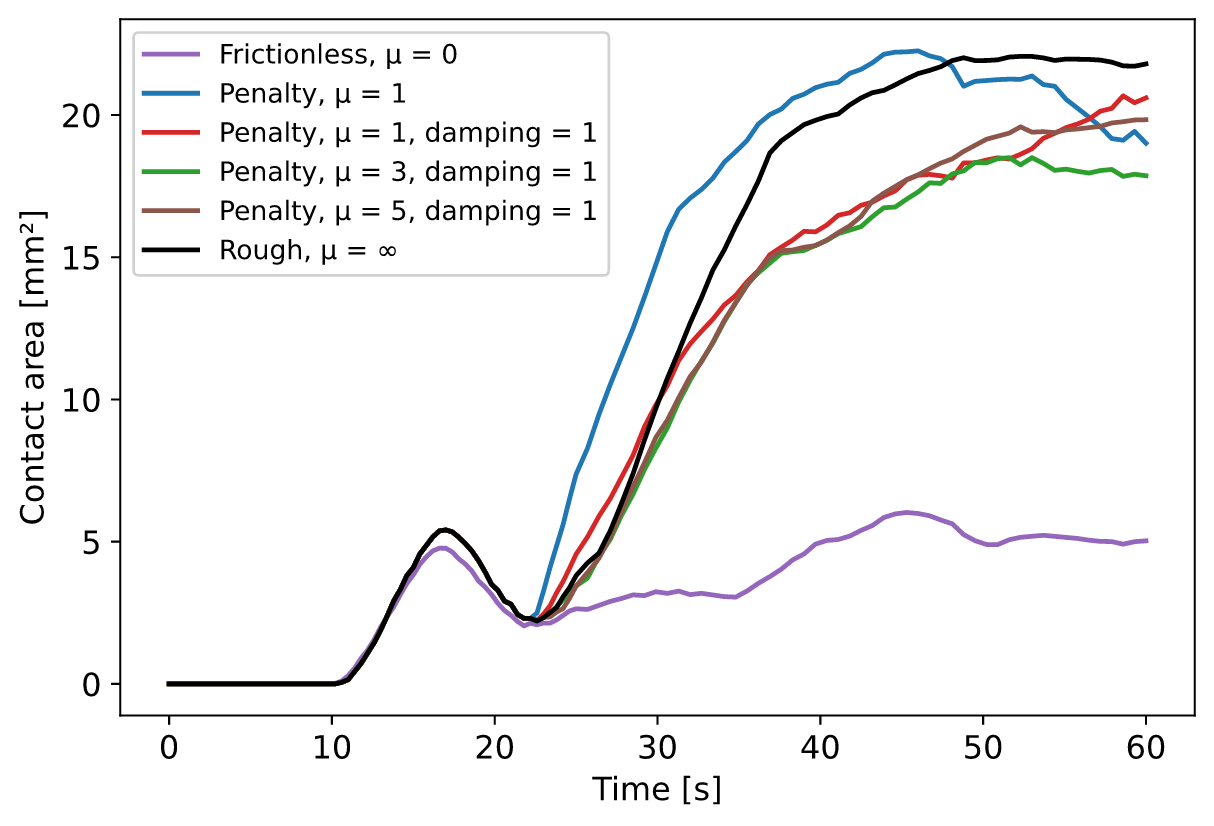}
\caption{Device-wall contact area.}
\label{fig:friction_contact_area}
\end{subfigure}
\hfill
\begin{subfigure}[t]{0.48\textwidth}
\centering
\includegraphics[width=\linewidth]{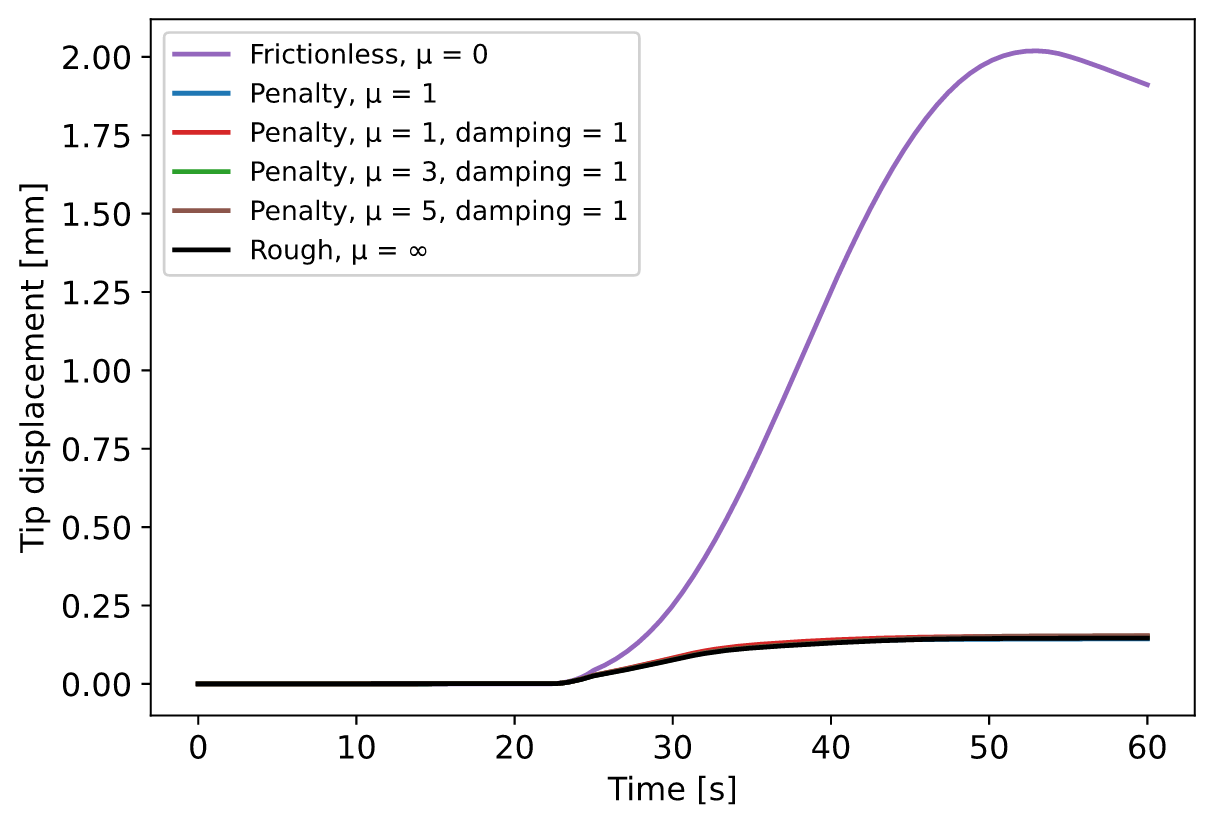}
\caption{Pole displacement.}
\label{fig:friction_tip_disp}
\end{subfigure}

\caption{Effect of device-wall tangential friction on the deployment response of Case~I. (a) Device-wall contact area and (b) monitored pole displacement are shown for the contact conditions defined in Table~\ref{tab:friction_cases}: \texttt{p0} ($\mu_{\mathrm{dw}}=0$), \texttt{p1} ($\mu_{\mathrm{dw}}=1$), \texttt{p1d1} ($\mu_{\mathrm{dw}}=1$ with contact damping), \texttt{p3d1} ($\mu_{\mathrm{dw}}=3$ with contact damping), \texttt{p5d1} ($\mu_{\mathrm{dw}}=5$ with contact damping), and \texttt{r} (rough contact). Frictionless wall contact permits pronounced late post-contact motion, whereas finite tangential friction strongly suppresses post-contact pole motion. The rough-contact case represents the limiting no-slip condition rather than a physical friction coefficient.}
\label{fig:friction_response}
\end{figure*}

\subsection{Reference deployment}
\label{subsec:baseline}

Case~I serves as the reference anatomy. The device opens from the crimped state into a sac-filling, ostium-oriented configuration. The pole remains near the neck region rather than moving deep into the dome, and distributed contact develops along the neck-adjacent wall. No immediate inversion, severe flattening, or deep sac-side migration is observed at the end of the simulated release. The final shape is not imposed by geometric morphing; it results from superelastic recovery, wire-wire contact, wire-wall contact, wall compliance, and tangential friction.

Table~\ref{tab:deployment_metrics} reports endpoint kinematic and neck-engagement metrics for the $z=0$ rough-contact limiting case. These values characterize one bounded contact condition and are not interpreted as physiological reference values.

\begin{table}[htbp!]
\centering
\small
\caption{Endpoint metrics for Case~I with rough device-wall contact and release depth $z=0$.}
\label{tab:deployment_metrics}
\begin{tabular}{l c}
\toprule
Metric & Value \\
\midrule
Final pole displacement, $u_{\mathrm{tip}}$ & $0.146~\mathrm{mm}$ \\
Final device-wall contact area, $A_{\mathrm{contact}}$ & $22.05~\mathrm{mm}^2$ \\
Maximum device-wall contact area & $24.02~\mathrm{mm}^2$ \\
Neck-rim coverage, $\eta_{\mathrm{cov}}$ & $85.7\%$ \\
Maximum uncovered rim gap, $g_{\max}$ & $2.21~\mathrm{mm}$ \\
Neck circumference, $L_{\mathrm{neck}}$ & $29.74~\mathrm{mm}$ \\
Equivalent neck area, $A_{\mathrm{neck,eq}}$ & $70.37~\mathrm{mm}^2$ \\
Coverage-equivalent neck area, $A_{\mathrm{cov,eq}}$ & $60.29~\mathrm{mm}^2$ \\
\bottomrule
\end{tabular}
\end{table}

\subsection{Sensitivity to device-wall tangential friction}
\label{subsec:contact_friction}

For the reference Case~I anatomy, the device-wall tangential condition is varied to examine its influence on the deployment trajectory and late-time wall engagement. Device-wall contact area and pole displacement quantify wall engagement and residual post-contact motion, respectively. Contact area remains small during the early constrained phase, increases as the braid expands against the wall, and then evolves toward a late-time plateau (Figure~\ref{fig:friction_contact_area}). Pole displacement follows the complementary kinematic response: little motion during the constrained phase, rapid movement during release, and either stabilization or continued post-contact motion (Figure~\ref{fig:friction_tip_disp}).

The sensitivity analysis reveals two distinct response regimes (Figure~\ref{fig:friction_response}). Under frictionless device-wall contact, normal wall engagement does not arrest tangential motion, and the device undergoes pronounced post-contact pole displacement despite the formation of substantial wall contact. Introducing finite tangential friction changes this response qualitatively: pole motion is strongly suppressed after wall engagement, while the contact area continues to evolve through local braid rearrangement. Increasing the resistance toward the rough-contact limit further restricts interfacial sliding, but does not produce a proportionally larger contacted area. Thus, wall engagement and kinematic stabilization are distinct mechanical quantities. Within the present sensitivity study, finite tangential friction is required to suppress global post-contact device motion while retaining local freedom for braid conformation.

\subsection{Sensitivity to release depth}
\label{subsec:release_depth}
The release-depth sensitivity analysis is performed for the reference Case~I anatomy. Release depth is an influential procedural parameter because it determines where the first strong wall reactions develop and whether those reactions guide the device toward the ostium or deeper into the sac. A deeper release tends to establish wall contact below the neck and can reinforce sac-side seating, whereas a neck-adjacent release can generate early ostial contact before the braid has fully expanded into the aneurysm. Both paths can yield plausible late-time configurations while differing substantially in their contact history.

\begin{figure*}[htbp!]
\centering

\begin{subfigure}[t]{0.48\textwidth}
\centering
\includegraphics[width=\linewidth]{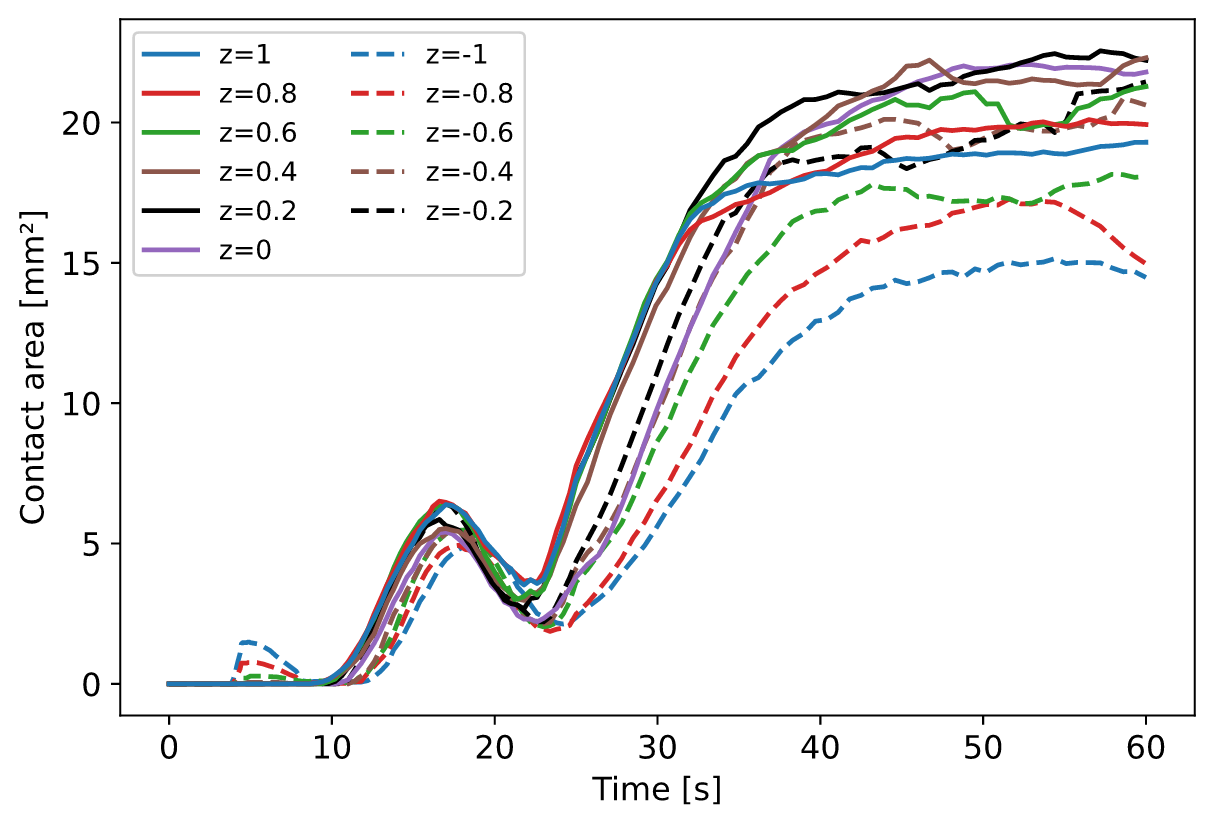}
\caption{Device-wall contact area.}
\label{fig:depth_contact_area}
\end{subfigure}
\hfill
\begin{subfigure}[t]{0.48\textwidth}
\centering
\includegraphics[width=\linewidth]{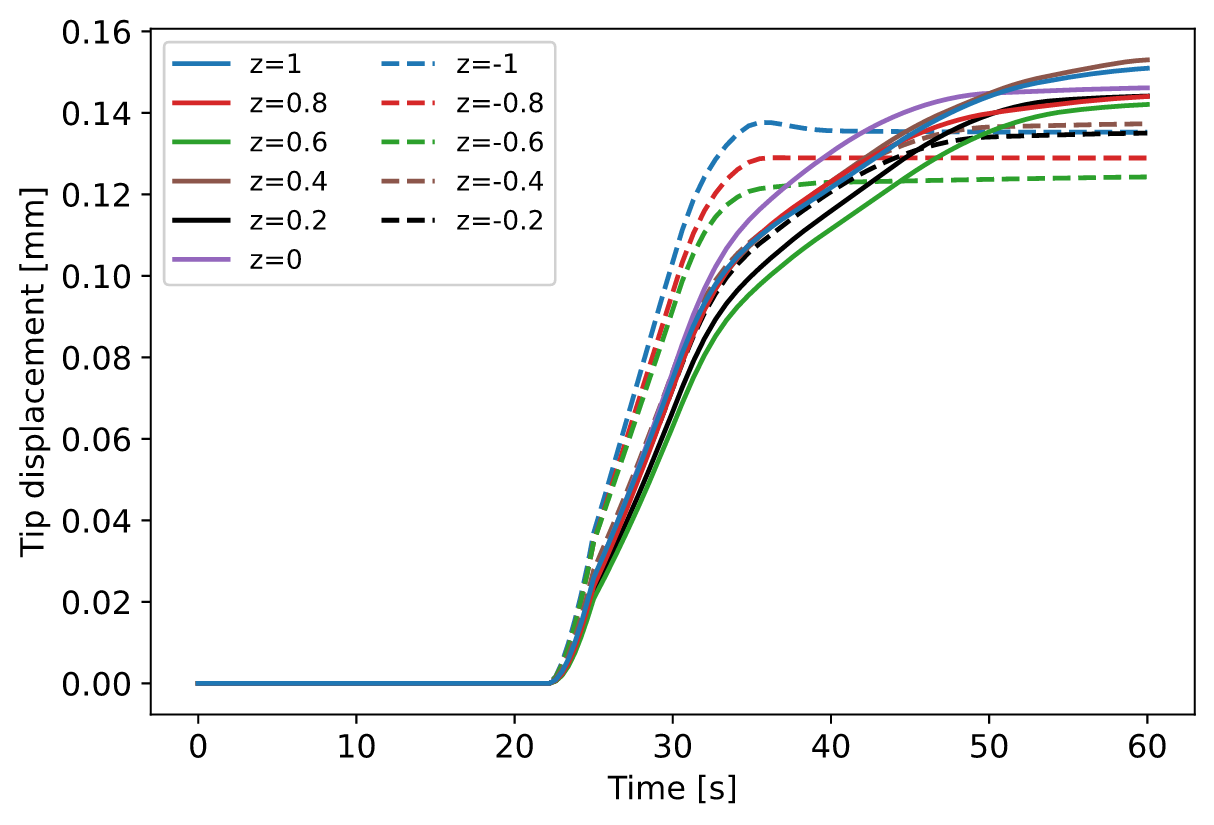}
\caption{Pole displacement.}
\label{fig:depth_tip_disp}
\end{subfigure}

\caption{Sensitivity of the Case~I deployment response to the prescribed initial release depth relative to the aneurysm neck plane. (a) Device-wall contact area and (b) monitored pole displacement. Release depth modifies both the onset and growth of wall engagement and the subsequent kinematic path of the device, demonstrating that similar late-time pole positions can arise from mechanically distinct deployment histories.}
\label{fig:release_depth_response}
\end{figure*}

The release-depth histories show changes in contact onset, contact-growth rate, and residual pole motion (Figures~\ref{fig:depth_contact_area} and \ref{fig:depth_tip_disp}). Release depth therefore changes the path to the late-time state, not only the final visual pose. The neck-rim metrics $\eta_{\mathrm{cov}}$ and $g_{\max}$ provide complementary measures of ostium engagement, whereas total device-wall contact area also includes contact elsewhere within the aneurysm sac. The corresponding endpoint values for the reported $z=0$ rough-contact case are listed in Table~\ref{tab:deployment_metrics}.

Ultimately, these parametric variations reveal that the final deployed state depends fundamentally on path history. Different interfacial friction levels and procedural insertion depths modify the initial contact onset, wall engagement kinetics, and residual pole motion.

\subsection{Mechanically resolved deployment and geometric reconstruction}
\label{subsec:fem_fast}

\begin{figure*}[t!]
\centering
\begin{subfigure}[b]{0.32\textwidth}
\centering
\adjustbox{trim=275pt 0pt 275pt 0pt, clip, width=0.82\textwidth}{%
\includegraphics{p8_cns_aneurysm_dispplot.png}
}
\caption{FEM, Case I}
\end{subfigure}
\hfill
\begin{subfigure}[b]{0.32\textwidth}
\centering
\adjustbox{trim=250pt 0pt 180pt 0pt, clip, width=0.86\textwidth}{%
\includegraphics{p4_cns_aneurysm_dispplot.png}
}
\caption{FEM, Case II}
\end{subfigure}
\hfill
\begin{subfigure}[b]{0.32\textwidth}
\centering
\adjustbox{trim=250pt 0pt 150pt 0pt, clip, width=0.86\textwidth}{%
\includegraphics{p6_cns_aneurysm_dispplot.png}
}
\caption{FEM, Case III}
\end{subfigure}

\vspace{0.6em}

\begin{subfigure}[b]{0.32\textwidth}
\centering
\includegraphics[width=0.95\textwidth]{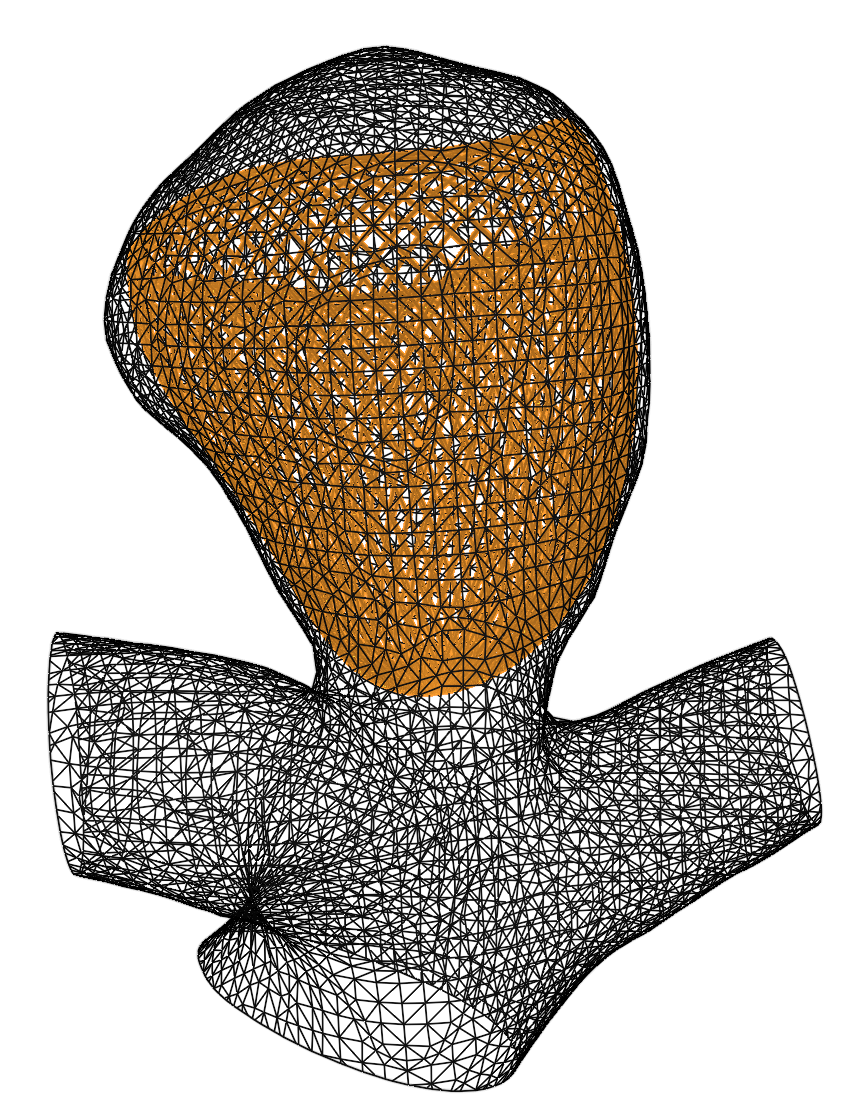}
\caption{Geometric reconstruction, Case I}
\end{subfigure}
\hfill
\begin{subfigure}[b]{0.32\textwidth}
\centering
\includegraphics[width=0.95\textwidth]{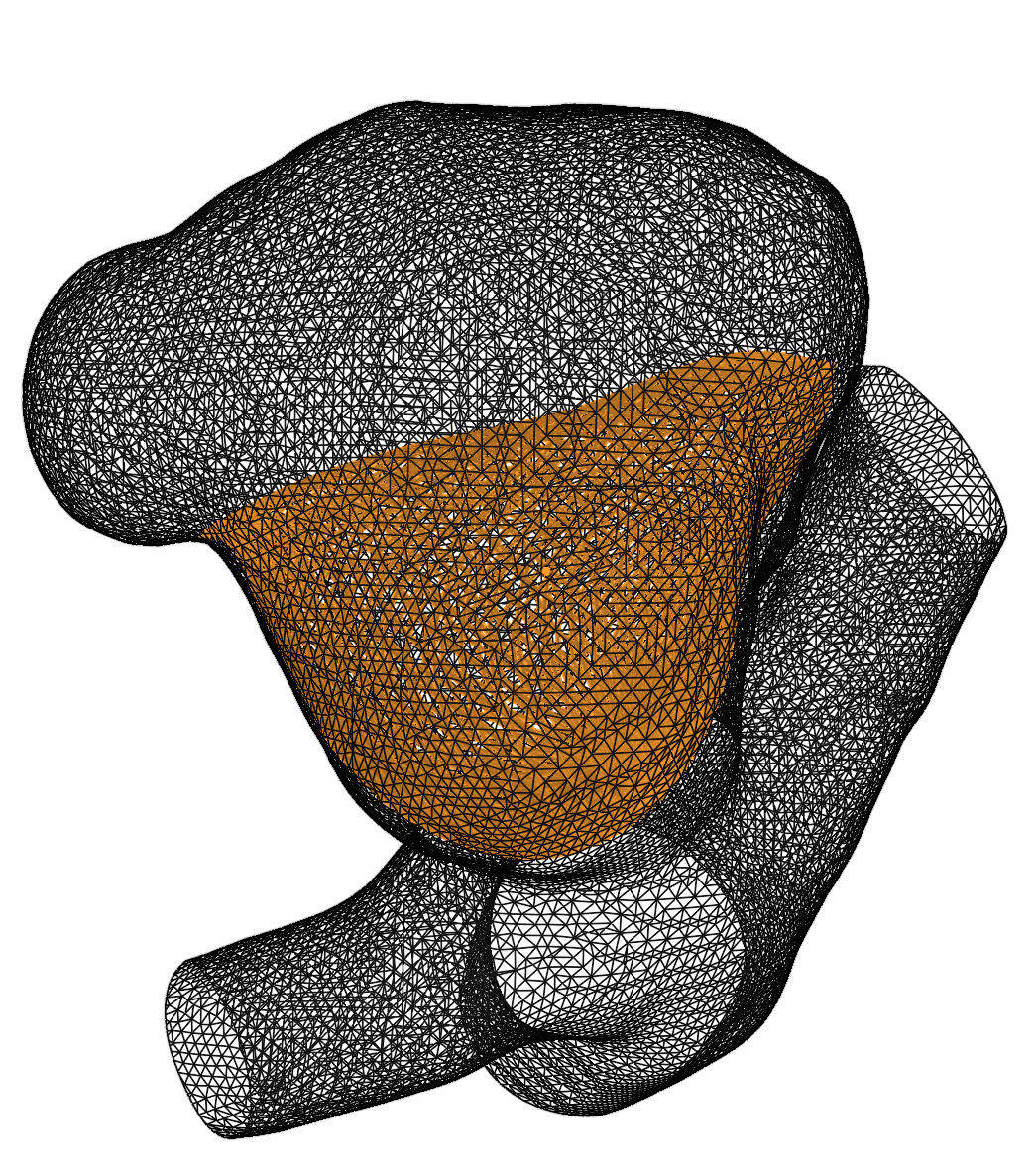}
\caption{Geometric reconstruction, Case II}
\end{subfigure}
\hfill
\begin{subfigure}[b]{0.32\textwidth}
\centering
\includegraphics[width=0.95\textwidth]{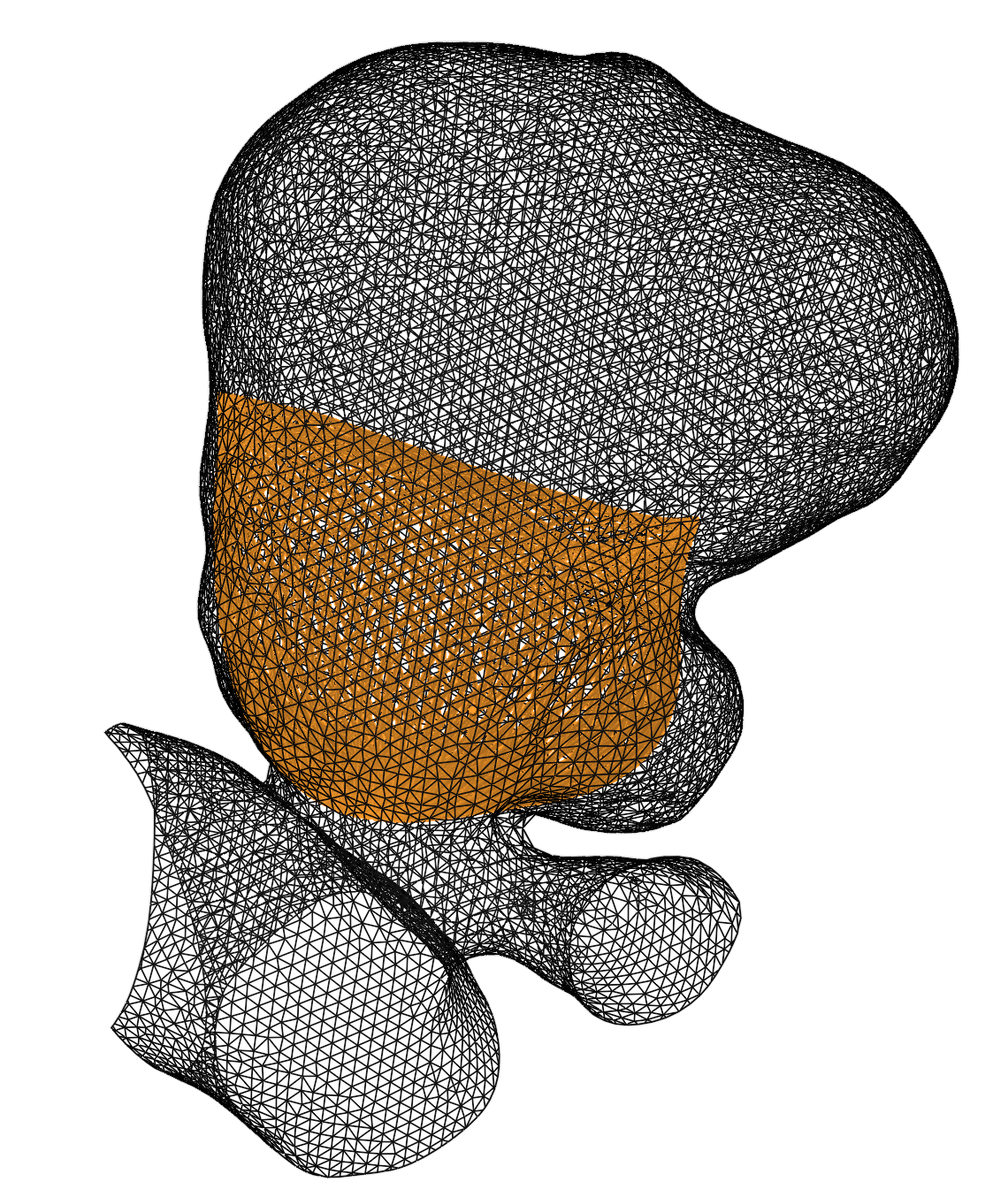}
\caption{Geometric reconstruction, Case III}
\end{subfigure}
\caption{Front-view comparison of contact-resolved Finite Element Method (FEM) deployment and imaging-guided geometric reconstruction for all three patient-specific anatomies. Panels are not displayed at a common physical scale and are intended for qualitative comparison of the deployed configurations. The geometric reconstructions provide qualitative image-based reference poses but do not resolve interface mechanics, frictional sliding, self-contact, contact tractions, or wall-supported state.}
\label{fig:fem_fast_all_cases}
\end{figure*}

Figure~\ref{fig:fem_fast_all_cases} contrasts the mechanically resolved finite-element deployments with the imaging-guided geometric reconstructions across all three patient-specific anatomies. While the manual geometric reconstructions provide a visually plausible spatial pose, they inherently omit underlying mechanical variables such as evolving wire-wire contact, localized braid rearrangement, localized contact area, and structural wall support. By contrast, the finite-element framework directly computes these mechanically resolved quantities, demonstrating that a visually matched geometry alone is insufficient to characterize the mechanically admissible, wall-supported state of the implanted device.

Crucially, as visualized in Figure~\ref{fig:fem_fast_all_cases}, our framework uncovers the localized structural deformation of the hyperelastic aneurysm wall under the expansion pressure of the braided Nitinol mesh. Because the biological tissue is modeled as a compliant, deformable domain rather than a rigid surface, the physical device-wall interaction drives an anatomical shape modification. This outward wall apposition and mechanical compliance represent a vital biophysical aspect of clinical translation that geometric reconstruction or simplified fast-placement strategies are inherently incapable of capturing without a high-fidelity, contact-resolved formulation.

\subsection{Transfer across anatomies}
\label{subsec:anatomy_transfer}

The same governing formulation, device representation, constitutive description, and contact treatment are applied to Cases~II and III without prescribing a common implanted device shape (Figure~\ref{fig:fem_fast_all_cases}). The resulting late-time configurations differ markedly among the three anatomies. Case~I produces a comparatively axial, ostium-oriented configuration, whereas Cases~II and III exhibit stronger asymmetric confinement and correspondingly different device orientations and neck-adjacent support patterns. These differences arise naturally from the patient-specific wall geometry and the associated contact history rather than from geometry-specific morphing of the device.

The three cases are not intended to establish statistical relationships between aneurysm morphology and deployment success. Rather, they demonstrate that a single wire-resolved contact formulation can generate distinct wall-supported configurations across substantially different anatomical environments. This transferability is a prerequisite for future cohort-based studies in which morphological descriptors can be related quantitatively to deployment outcomes.

\section{Discussion}
\label{sec:discussion}

The present simulations demonstrate that CNS deployment must be treated as a path-dependent contact problem when the aim is to resolve the mechanically seated configuration. The implanted state cannot be inferred from nominal device diameter or from a visually fitted final shape. It emerges from the interaction of mechanical braid behavior, wire-wire contact, wall contact, tangential slip, wall compliance, and release history. Once this interaction is solved, pole position, rim coverage, and uncovered neck gaps become mechanical outputs rather than assumptions.

This framework provides a deterministic means to resolve the mechanical contact history, such as evolving wire-wire friction, braid expansion kinetics, and the mechanically resolved wall-supported state, that is not resolved by standard \textit{a priori} virtual placement strategies. The present results do not establish that the computed configuration is a quantitatively validated representation of the in vivo device state. Rather, they provide a framework in which these quantities can be computed and subsequently validated.

Device-wall tangential friction governs post-contact sliding and anchoring against the aneurysm wall, whereas wire-wire friction is kept fixed at $\mu_{\mathrm{ww}}=0.01$ to permit braid rearrangement. The present simulations are designed to examine the structural consequences of bounded tangential slip resistance, not to estimate a physiological or clinically transferable coefficient of friction. Because the wire-wire and wire-wall interfaces are not calibrated against device-specific, blood-wetted bench measurements, the prescribed device-wall friction levels should be interpreted as contact regimes used to quantify the mechanical consequences of tangential slip resistance, rather than as calibrated physiological interface properties. In particular, the $\mu=3$, $\mu=5$, and rough-contact cases are upper-bound calculations rather than candidate values for patient-specific prediction.

Predictive application requires separate experimental calibration of the wire-wire and wire-wall interfaces using the actual device surface finish, compliant vascular tissue or representative phantoms, and a blood- or saline-wetted environment. Until such calibration is available, frictional uncertainty should be propagated through a bounded sensitivity analysis rather than represented by a single nominal value. At wire-wire crossovers, the fixed low coefficient $\mu_{\mathrm{ww}}=0.01$ permits slip, compaction, and local pore redistribution. At the device-wall interface, the varied coefficient $\mu_{\mathrm{dw}}$ controls post-contact sliding and anchoring against the aneurysm wall. Low tangential friction allows the device to adapt but can permit residual post-contact motion. A near-stick limit anchors the device strongly but may suppress useful rearrangement. This trade-off shows why a frictional interface model is needed to investigate deployment sensitivity, even though the present friction coefficients are not yet physiologically calibrated. A purely geometric placement cannot represent this contact-dependent sensitivity.

Release depth is equally important because it sets the location of early wall engagement and, therefore, the direction of the reactions that guide expansion. For a neck-covering device, the clinically relevant endpoint is not generic sac occupancy but stable ostium engagement. Thus, vertical placement should be treated as a procedural parameter in virtual planning, not as a minor geometric adjustment. The prescribed release duration is kept fixed in the present study; only the device-wall friction condition and initial release depth are varied. Although the constitutive response is formulated as rate-independent, the computed deployment path in an explicit analysis may remain sensitive to inertial effects, contact regularization, and numerical damping. Future studies can vary the release duration while monitoring kinetic-to-internal energy ratios, frictional dissipation, contact-area evolution, and pole-stabilization time to distinguish mechanically driven tangential friction from numerical dissipation.

The advantage of the contact-resolved FEM approach is that it computes a wall-supported deployed configuration together with wire-wire contact, device-wall contact, tangential slip, and contact tractions. Imaging-guided geometric reconstruction remains useful for rapid visualization and comparison with post-treatment imaging, but it does not provide these mechanical quantities. Therefore, it may be sufficient for exploratory fixed-geometry flow calculations, whereas mechanics-based deployment is required when the question concerns deployment sensitivity, device-wall interaction, wall loading, migration tendency, or FSI. Further workflow automation and computational acceleration are required before intra-procedural application. This distinction is also relevant for subsequent mechanobiological modeling because the deployment-generated pore geometry, neck apposition, and uncovered inflow pathways define the mechanical and hemodynamic initial conditions from which post-treatment flow alteration, residence-time changes, thrombus formation, and longer-term device-tissue interaction evolve.

The next validation step is a controlled bench-deployment study in patient-specific or idealized aneurysm phantoms, with measured device shapes, pole positions, and neck-level apposition used to calibrate the Nitinol response, frictional contact parameters, and wall compliance. After that, larger retrospective cohorts with high-resolution post-treatment imaging, including longitudinal assessment of aneurysm and device morphology \cite{spitz2026morphological}, can test whether the simulated seated states reproduce observed device pose, deformation, and migration tendencies. Once this structural validation is in place, the contact-consistent deployed geometries serve as input for hemodynamic simulations.

Despite the structural insights provided by this contact-resolved framework, several limitations remain as key avenues for future investigation. First, the predictive capacity of the current deployment model remains bounded by the use of homogeneous surrogate material properties for both the hyperelastic aneurysm wall and the superelastic Nitinol wires, omitting patient-specific tissue variability and device-specific thermomechanical calibrations. Second, the wire-wire and device-wall friction coefficients utilized herein represent idealized mechanical regimes; accurate \textit{in vivo} prediction demands rigorous experimental calibration using physical samples within blood- or saline-wetted anatomical phantoms. Additionally, while the three patient-specific cases successfully demonstrate the multi-stage computational workflow, a significantly larger clinical cohort is required to establish robust statistical correlations between diverse aneurysm morphologies and deployment outcomes. Finally, this study focuses strictly on the isolated solid mechanics of the implant; translating these contact-consistent configurations into FSI pipelines, alongside tracking deployment-informed thrombus formation, constitutes the immediate next endeavour of this research framework.

\section{Conclusions}
\label{sec:conclusions}

This study presents the foundations of a patient-specific FEM framework for contact-resolved deployment of the CNS. The framework represents the device as a wire-resolved dual-layer braid, couples it to a deformable aneurysm wall, and computes the seated state through frictional contact and staged release. In the reference anatomy, device-wall tangential friction and release depth strongly influenced ostium-centered seating, wall engagement, and late-time device motion. The mechanics-based approach resolves interface contact, tangential sliding, contact tractions, and the resulting wall-supported configuration, quantities that are not available from purely geometric placement. The framework provides a basis for generating deployment-informed geometries for downstream hemodynamic and FSI studies. Structural validation and device-specific calibration remain necessary before predictive patient-specific application.

The immediate next step is to use the deployed, contact-consistent geometries as input for hemodynamic simulations. Future work extends the framework toward one-way and fully coupled FSI \cite{hagmeyer2021fluid,hagmeyer2022one,hagmeyer2024fully}, so that wall motion, device motion, and flow alteration can be studied together. A further objective is thrombus-formation modeling, including deployment-informed residence time, transport, and clot-growth indicators at the aneurysm neck and within the sac. Larger patient cohorts, bench validation, and device-specific Nitinol calibration are needed before clinical decision support is attempted.

\section*{Acknowledgments}
The authors thank Cerus Endovascular for providing a physical sample of the device, which facilitates inspection of its braided architecture.

\section*{Statements and Declarations}

\subsection*{Funding}
This research is supported by the IDIR-Project (Digital Implant Research), a cooperation financed by Kiel University, University Hospital Schleswig-Holstein, and Helmholtz Zentrum Hereon. M.F. and A.P. acknowledge funding by the Deutsche Forschungsgemeinschaft (DFG, German Research Foundation), project number 465242983. R.P. acknowledges support for open-access publication from the Universität der Bundeswehr München under Projekt DEAL.

\subsection*{Competing interests}
The authors declare no competing interests.

\subsection*{Author contributions}
R.P.: conceptualization, methodology, software, formal analysis, visualization, writing -- original draft, writing -- review and editing. F.G.: software, writing -- review and editing. M.F., I.S., M.M., S.S., and N.L.: writing -- review and editing. A.P.: conceptualization, funding acquisition, writing -- review and editing. All authors read and approve the final manuscript.

\subsection*{Data availability}
The geometry, input files, and post-processing scripts underlying the present computational study are available from the corresponding author upon reasonable request.

\bibliographystyle{elsarticle-num}
\bibliography{example}

\end{document}